\documentclass[twocolumn,floatfix,prx,aps,showpacs,longbibliography]{revtex4-2}
\usepackage{graphicx,amsmath,amssymb,color}
\usepackage{nicefrac}
\usepackage[titletoc,title]{appendix}
\usepackage[colorlinks,bookmarks=true,citecolor=blue,linkcolor=red,urlcolor=blue]{hyperref}

\newcommand{\be}{\begin{equation}}
\newcommand{\ee}{\end{equation}}

\newcommand{\ba}{\begin{eqnarray}}
\newcommand{\ea}{\end{eqnarray}}

\begin{document}

\title{Transitions from Abelian composite fermion to non-Abelian parton fractional quantum Hall states in the zeroth Landau level of bilayer graphene}
\author{Ajit C. Balram}
\affiliation{Institute of Mathematical Sciences, CIT Campus, Chennai 600113, India}
\affiliation{Homi Bhabha National Institute, Training School Complex, Anushaktinagar, Mumbai 400094, India}
\date{\today}

\begin{abstract} 
The electron-electron interaction in the Landau levels of bilayer graphene is markedly different from that of conventional semiconductors such as GaAs. We show that in the zeroth Landau level of bilayer graphene, in the orbital which is dominated by the nonrelativistic second Landau level wave function, by tuning the magnetic field a topological quantum phase transition from an Abelian composite fermion to a non-Abelian parton fractional quantum Hall state can be induced at filling factors $1/2, ~2/5$, and $3/7$.  The parton states host exotic anyons that can potentially be utilized to store and process quantum information.  Intriguingly, some of these transitions may have been observed in a recent experiment [K. Huang \emph{et al.} arXiv:2105.07058].
\end{abstract}

\maketitle
Traditionally, semiconductor quantum wells such as those in GaAs/AlGaAs have been the system of choice to experimentally study fractional quantum Hall effect (FQHE) physics~\cite{Tsui82, Willett87}. Graphene, with its relativistic dispersion and the presence of multiple components such as spins, valleys, orbitals, and layers adds to the richness of the FQHE phenomenology~\cite{Neto09, Xu09, Bolotin09, Dean11, Feldman12, Feldman13, Amet15, Kim19}. Aside from these features, in multilayer graphene systems the interactions between electrons can be controlled by parameters such as perpendicular magnetic and electric fields which can assist in stabilizing exotic FQHE states.

Robust even-denominator FQHE states with gaps of the order of a few degrees Kelvin have been observed in the zeroth Landau level (ZLL) of Bernal-stacked bilayer graphene (BLG)\cite{Zibrov16, Li17}. When the LL with $\mathcal{N}{=}0$ orbitals is partially filled, the Jain sequence of odd-denominator Abelian FQHE states described in terms of composite fermions (CFs)~\cite{Jain89} is seen. On the other hand, in the LL with $\mathcal{N}{=}1$ orbitals, only the states at filling factors $\nu=1/3$, $2/3$, and $1/2$ are well established while at some other fractions signatures of FQHE were observed. More recently, Huang~\emph{et al.}~\cite{Huang21} have observed extensive FQHE in the ZLL of BLG. Furthermore, they showed that transitions between FQHE states at $\nu=2/5,~3/7$, and $1/2$ can be induced by varying the magnetic field or applying an electric field. The primary result of our work is to show that the transitions they observed are likely from Abelian CF to non-Abelian ``parton'' states. Encouragingly, these results suggest that BLG could potentially serve as a platform to host Fibonacci anyons which can perhaps form the building blocks of a universal fault-tolerant quantum computer.

\underline{\emph{Zeroth Landau level of bilayer graphene.}} The zero-energy manifold in BLG has eight LLs in it with two each coming from the spin ($|{\uparrow}\rangle,|{\downarrow}\rangle$), valley ($|{+}\rangle,|{-}\rangle$), and orbital $(\mathcal{N}{=}0,1)$ degrees of freedom~\cite{Shibata09}. The ordering of these single-particle states and their orbital character can be varied by an interlayer electric field and a magnetic field, respectively~\cite{Zibrov17, Huang21, Apalkov11, Papic11, Hunt17, Zhu20a}. The $\mathcal{N}{=}0$ LLs are identical to the $n{=}0$ LL [lowest LL (LLL)] of GaAs ($n$ denotes the LL index of conventional semiconductors while $\mathcal{N}$ refers to the LL index for graphene). However, the $\mathcal{N}{=}1$ LLs have an admixture of $n{=}0$ and $n{=}1$ [second LL (SLL)] orbitals; at small (large) magnetic fields, their orbital nature is more $n{=}1$ ($n{=}0$) like.  

We model the single-particle spinor wave function for the $\mathcal{N}{=}1$ LLs as [$\sin(\theta)\phi_{1}$,$\cos(\theta)\phi_{0}$]~\cite{Apalkov11} where $\phi_{n}$ is the wave function of a nonrelativistic electron in the LL indexed by $n$ and $\theta$ is a tunable parameter called the mixing angle. The mixing angle is related to the magnetic field $B$ as $\tan(\theta){=}t\ell/(\sqrt{2}\hbar v_{F})$, where $t$ is the hopping integral (estimated to be ${\approx}$350 meV from calculations at zero magnetic field~\cite{Jung14}), $v_{F}$ is the Fermi velocity (typically $10^6$ m/s in graphene), and $\ell{=}\sqrt{\hbar c/(eB)}$ is the magnetic length. There are three special values of $\theta$ that are of particular interest: (a) $\theta{=}0$ corresponds to the LLL,  (b) $\theta{=}\pi/4$ corresponds to the first excited $\mathcal{N}{=}1$ LL of monolayer graphene (MLG1)~\cite{Balram15c}, and (c) $\theta{=}\pi/2$ corresponds to the SLL of GaAs. Therefore, for FQHE physics, this simplified model suffices to cover all the eight LLs since $\theta{=}0$ recovers the $\mathcal{N}{=}0$ LLs. 

Throughout this Letter, we shall neglect the effects of screening by gates, rotation between the layers, valley-symmetry breaking, and disorder.  We also neglect the effects of LL mixing and thus states related by particle-hole symmetry are considered on an equal footing. Furthermore, we shall restrict ourselves to only one-component states. In the case of two components, where the components can be considered as spins residing in the $n{=}0$ LL [therefore the interaction is SU(2) invariant], the spin-phase diagram of many FQHE states has been studied in detail in the past~\cite{Wu93, Park98, Jain07, Liu14, Balram15, Balram15a, Balram15c, Balram17}.  Recently, a detailed phase diagram of two-component states in the $n{=}0$ LL of double-layer graphene [two graphene layers separated by an insulator, such as hBN, which breaks the SU(2) symmetry of the interaction] has been worked out both experimentally~\cite{Liu19, Li19} and theoretically~\cite{Faugno20}. Under appropriate settings, these two-component states could also be stabilized in BLG. We leave out an exploration of multi-component FQHE states in BLG for the future.  Our attention will be solely focused on the single-component FQHE states that could arise in any of the LLs with $\mathcal{N}{=}1$ (denoted by the pseudospins $|1,\uparrow,\pm\rangle$ and $|1,\downarrow,\pm\rangle$). We refer to any of these LLs with $\mathcal{N}{=}1$ as the ZLL of BLG.

\underline{\emph{Parton states:}} The parton theory~\cite{Jain89b} was introduced by Jain as a generalization of his CF theory. In the parton theory,  one imagines dividing the electron into $q$ fictitious entities called partons. To construct a gapped state of the electrons, each of the partons is placed in an integer quantum Hall effect (IQHE) state at filling $n_\alpha$, where $\alpha{=}1,2,{\cdots}, q$ labels the various species of the partons. The resulting electronic state, denoted as ``$n_1n_2n_3{\cdots}$," is described by the wave function
\begin{equation}
\Psi^{n_1n_2n_3\cdots}_\nu = \mathcal{P}_{\rm LLL} \prod_{\alpha=1}^{q}\Phi_{n_\alpha}(\{z_j\}),
\label{eq:parton_wf}
\end{equation}
where the coordinate of the $j$th electron is given by the complex number $z_{j}{=}x_{j}{-}iy_{j}$, $\Phi_n$ is the Slater determinant wave function for $n$-filled LLs of nonrelativistic electrons, and $\mathcal{P}_{\rm LLL}$ denotes projection into the LLL. We allow the parton fillings to be negative, which we denote by $\bar{n}$, with $\Phi_{\bar{n}}{=}\Phi_{-n}{=}\Phi_n^*$.  In these states, the partons see a magnetic field that is anti-parallel to that seen by the electrons.  The parton theory can also capture compressible states. In particular, when $n{\rightarrow}\infty$, the wave function $\Phi_{n}$ describes the gapless Fermi sea (FS). 

As the partons are unphysical objects they have to be glued back together to recover the physical electrons. This gluing procedure is already implemented in the wave function given in Eq.~\eqref{eq:parton_wf} since the different parton species coordinates $z_j^\alpha$ are all set equal to the electron coordinate $z_j$, i.e.,$z_j^\alpha{=}z_j$ for all $\alpha$.  Each $\Phi_{n_\alpha}$ in Eq.~\eqref{eq:parton_wf} is made up of \emph{all} the electronic coordinates $\{z_{j}\}$. The density of each parton species is the same as the electronic density and all the partons see the same magnetic field that the electrons experience. Thus, the charge of the $\alpha$ parton species $e_\alpha{=}{-}e\nu  / n_\alpha$, where ${-}e$ is the electronic charge. The parton charges add up to that of the electron, which results in the constraint $\nu {=}[\sum_{\alpha{=}1}^{q} n_\alpha^{{-}1}]^{{-}1}$. A parton state with a repeated factor of $n$, with $|n|{\geq 2}$, hosts excitations that carry non-Abelian braiding statistics~\cite{Wen91}.

The $\nu{=}1/r$ Laughlin state~\cite{Laughlin83} , described by the wave function $\Psi_{\nu{=}1/r}^{\rm Laughlin}{=}\Phi^{r}_{1}$, can be re-interpreted as the $r$-parton state where all the partons form a $\nu{=}1$ IQHE state. The $\nu{=}s/(2ps{\pm}1)$ Jain/CF state, described by the wave function $\Psi_{\nu{=}s/(2ps\pm 1)}^{\rm Jain}{=}\mathcal{P}_{\rm LLL}\Phi^{2p}_{1}\Phi_{\pm s}$, can be viewed as a $(2p{+}1)$-parton state where one parton forms a $\nu{=}{\pm}s$ IQHE state and rest of the $2p$ partons form a $\nu{=}1$ IQHE state. The Rezayi-Read~\cite{Rezayi94} wave function for the CF Fermi sea (CFFS) at $\nu{=}1/2$ can be interpreted as a ``${\rm FS}11$" state, where one parton forms a Fermi sea and two partons form a $\nu{=}1$ IQHE state. Several parton states, beyond the Abelian Laughlin and Jain states, have been proposed as feasible candidates to describe FQHE plateaus that arise in the LLL~\cite{Balram21a, Balram21c, Dora22}, SLL~\cite{Balram18, Balram18a, Balram19, Balram20, Balram20b, Balram21,Faugno20b}, LLL of wide quantum wells~\cite{Faugno19}, and in the LLs of graphene~\cite{Wu17, Kim19, Faugno20a, Balram20, Balram21, Faugno20b}. Furthermore, recently some very high-energy excited states have also been described in terms of partons~\cite{Balram21d}. 

Motivated by a recent experiment~\cite{Huang21} we consider FQHE at $\nu{=}2/5,~3/7$, and $1/2$ in the ZLL of BLG. The parton states that are relevant to these fillings are: (a) $\nu{=}2/5$: (a1) $211$, and (a2) $\bar{2}^{3}1^{4}$~\cite{Balram19}, which lies in the same universality class as the particle-hole conjugate of the three-cluster Read-Rezayi state~\cite{Read99} which supports Fibonacci anyons, (b) $\nu{=}3/7$: (b1) $311$, and (b2) $\bar{3}^{2}1^{3}$~\cite{Faugno20a}, whose excitations, such as those of the $\bar{2}^{3}1^{4}$, are also parafermionic; and (c) $\nu{=}1/2$: (c1) FS$11$, and (c2) $\bar{2}^{2}1^{3}$~\cite{Balram18}, which lies in the same topological phase as the anti-Pfaffian state~\cite{Levin07, Lee07}. The $\bar{2}^{2}1^{3}$ state can be interpreted as a topological $p$-wave superconductor of CFs~\cite{Read00, Balram18}. The aforementioned noninteracting CF states are known to be stabilized in the LLL~\cite{Jain07} and MLG1~\cite{Amet15, Kim19} while the non-Abelian parton states likely prevail in the SLL~\cite{Balram18, Balram19, Faugno20}. In the SLL of GaAs, FQHE has been observed at $2/5$ and $1/2$~\cite{Willett87, Xia04, Pan08, Choi08, Kumar10, Zhang12} and some signatures of it have been seen at $3/7$~\cite{Choi08}.

\underline{\emph{Numerical results.}} All our calculations are carried out on the Haldane sphere~\cite{Haldane83}. In this geometry, $N$ electrons move on the spherical surface in the presence of a radial magnetic flux of $2Qhc/e$ ($2Q$ is an integer) which is generated by a magnetic monopole placed at the center of the sphere. In the LL indexed by $\mathcal{N}$, the total number of single-particle orbitals is $2l+1$, where $l{=}Q+\mathcal{N}$ is the shell-angular momentum. FQHE ground states on the sphere occur when $2l{=}N/\nu{-}\mathcal{S}$, where $\mathcal{S}$ is a rational number called the shift~\cite{Wen92}. The shift can often differentiate between candidate states occurring at the same filling. The shift of the parton state described by the wave function of Eq.~\eqref{eq:parton_wf} is $\mathcal{S}^{n_1n_2{\cdots}n_{q}}{=}\sum_{\alpha{=}1}^{q} n_\alpha$. Due to the rotational symmetry, the total orbital angular momentum $L$ and its $z$ component are good quantum numbers on the sphere. FQHE ground states are uniform, i.e., they have $L{=}0$ while excitations generically have $L{>}0$. Although the sphere is not the best geometry to study the gapless CFFS, in this work we will consider filled-shell CF states on the sphere that have previously been shown to serve as representatives of the uniform CFFS~\cite{Rezayi94, Balram15b, Balram17, Liu20}.

\begin{figure*}[htpb]
			\begin{center}			
				\includegraphics[width=0.329\textwidth]{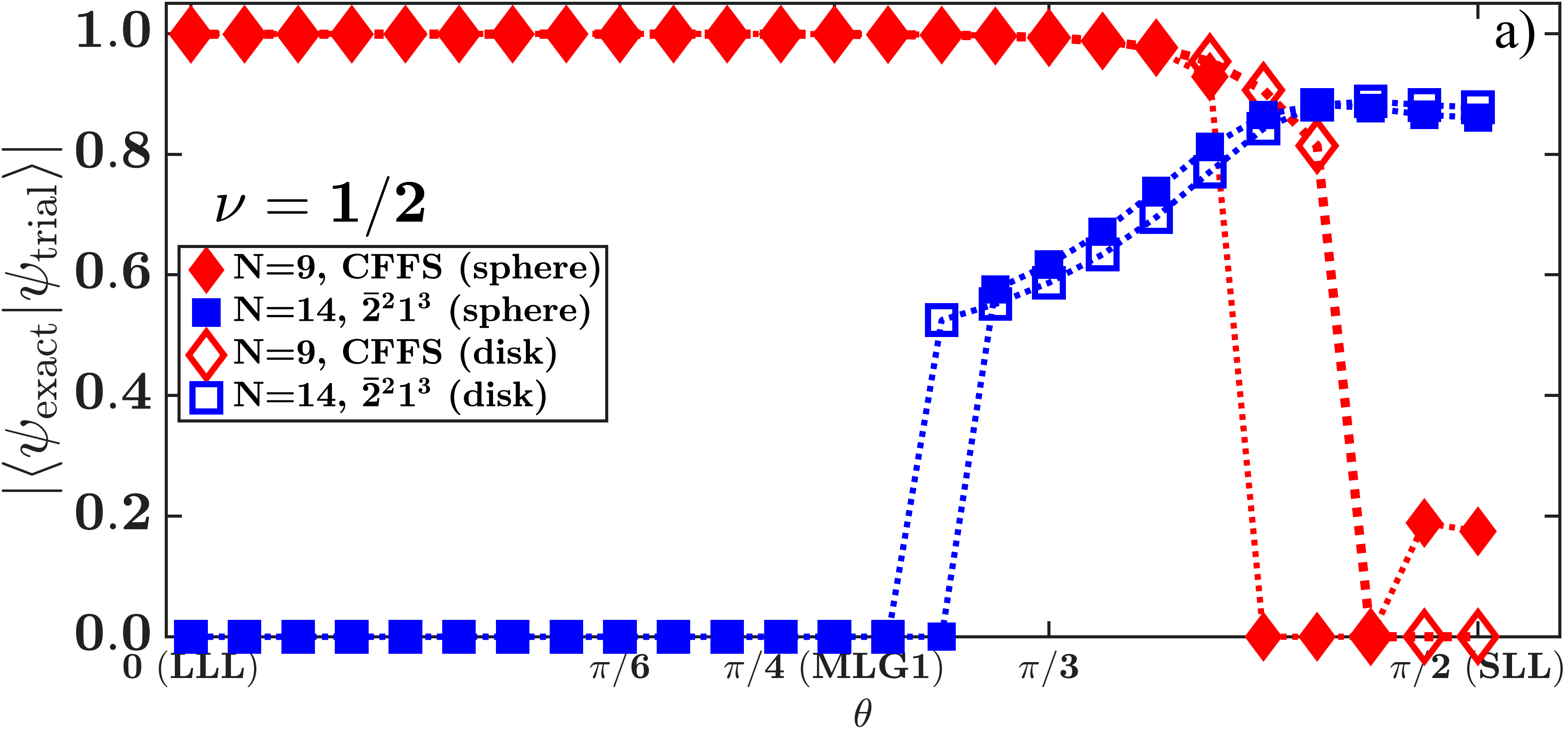}
				\includegraphics[width=0.329\textwidth]{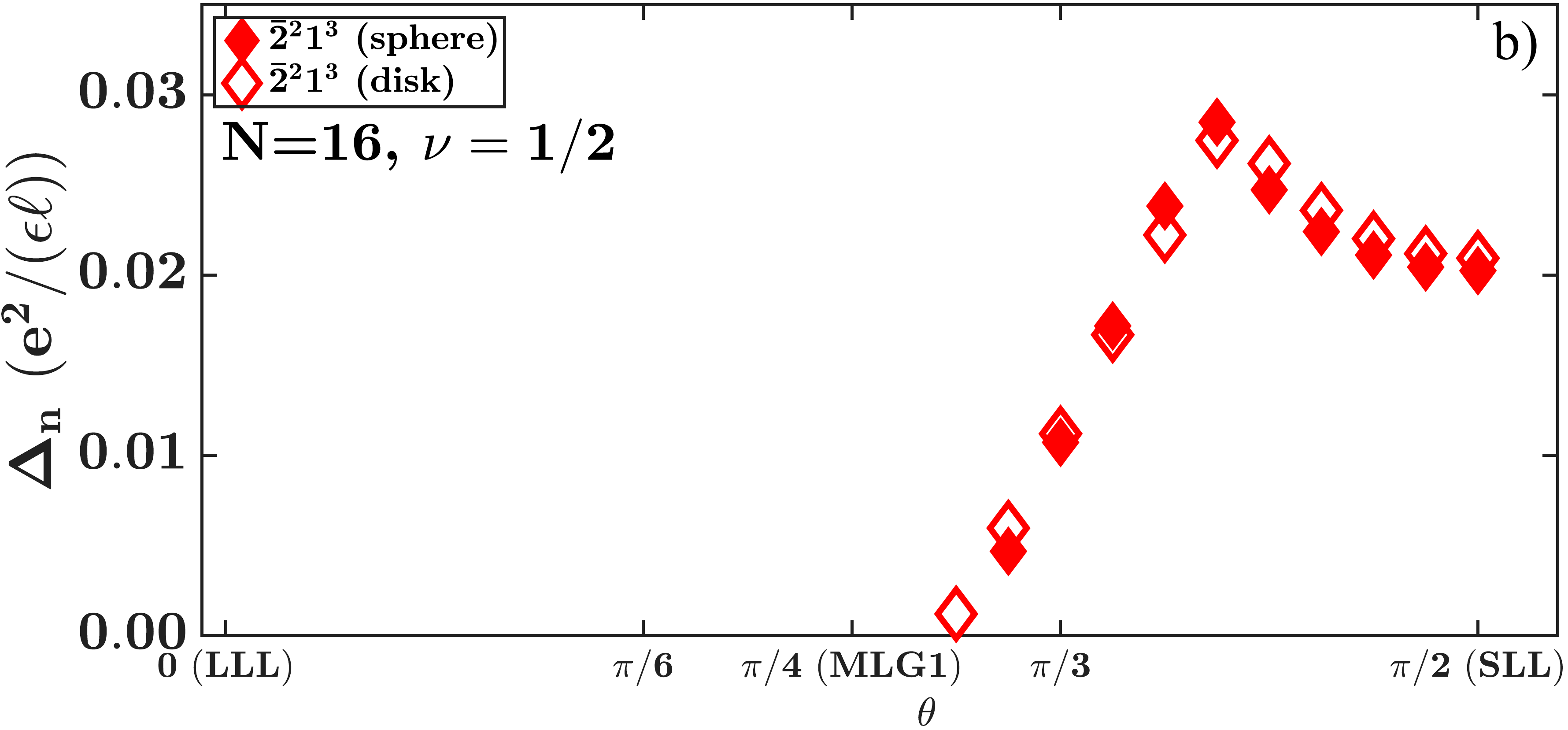}
				\includegraphics[width=0.329\textwidth]{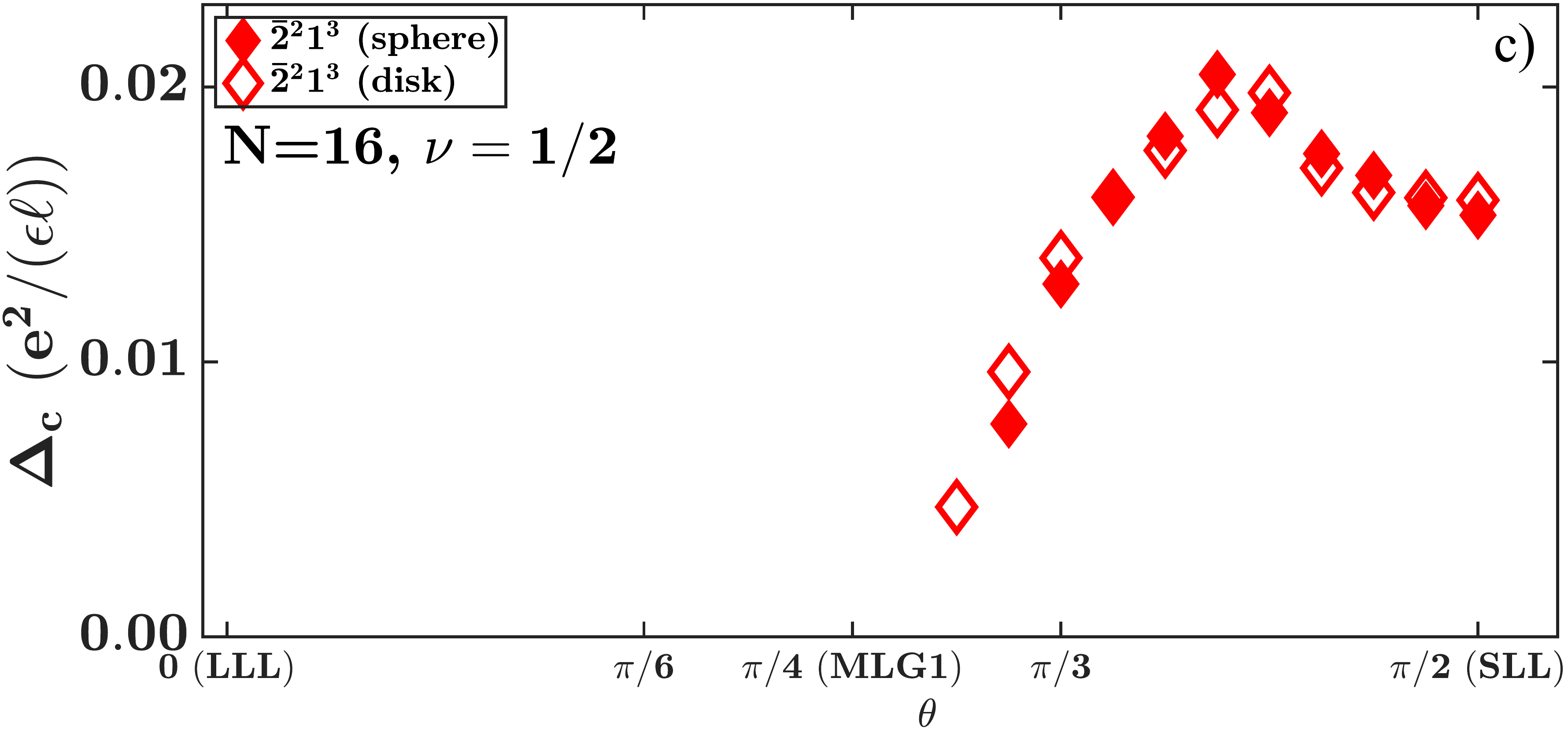}\\
				\vspace{0.3cm}
				\includegraphics[width=0.329\textwidth]{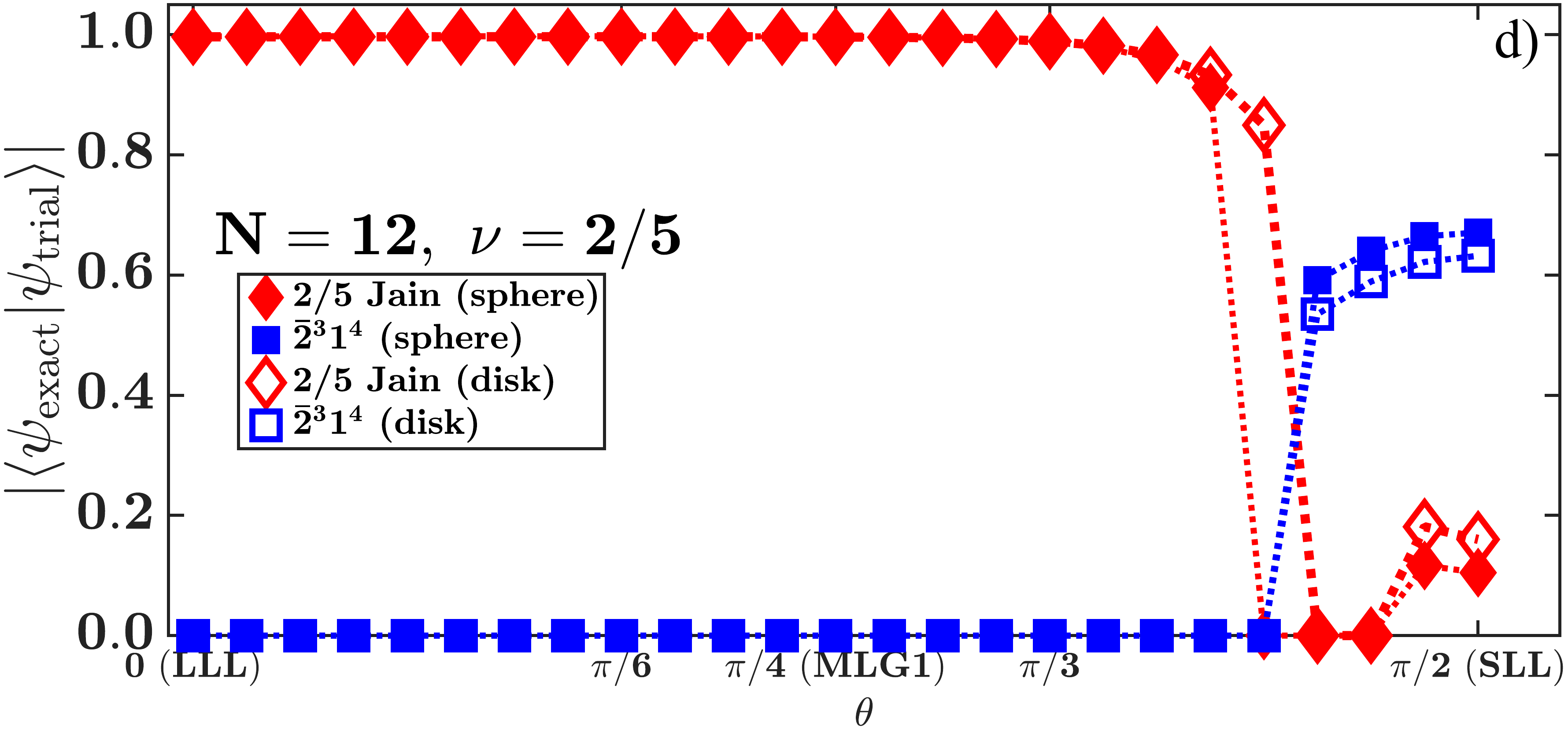}
				\includegraphics[width=0.329\textwidth]{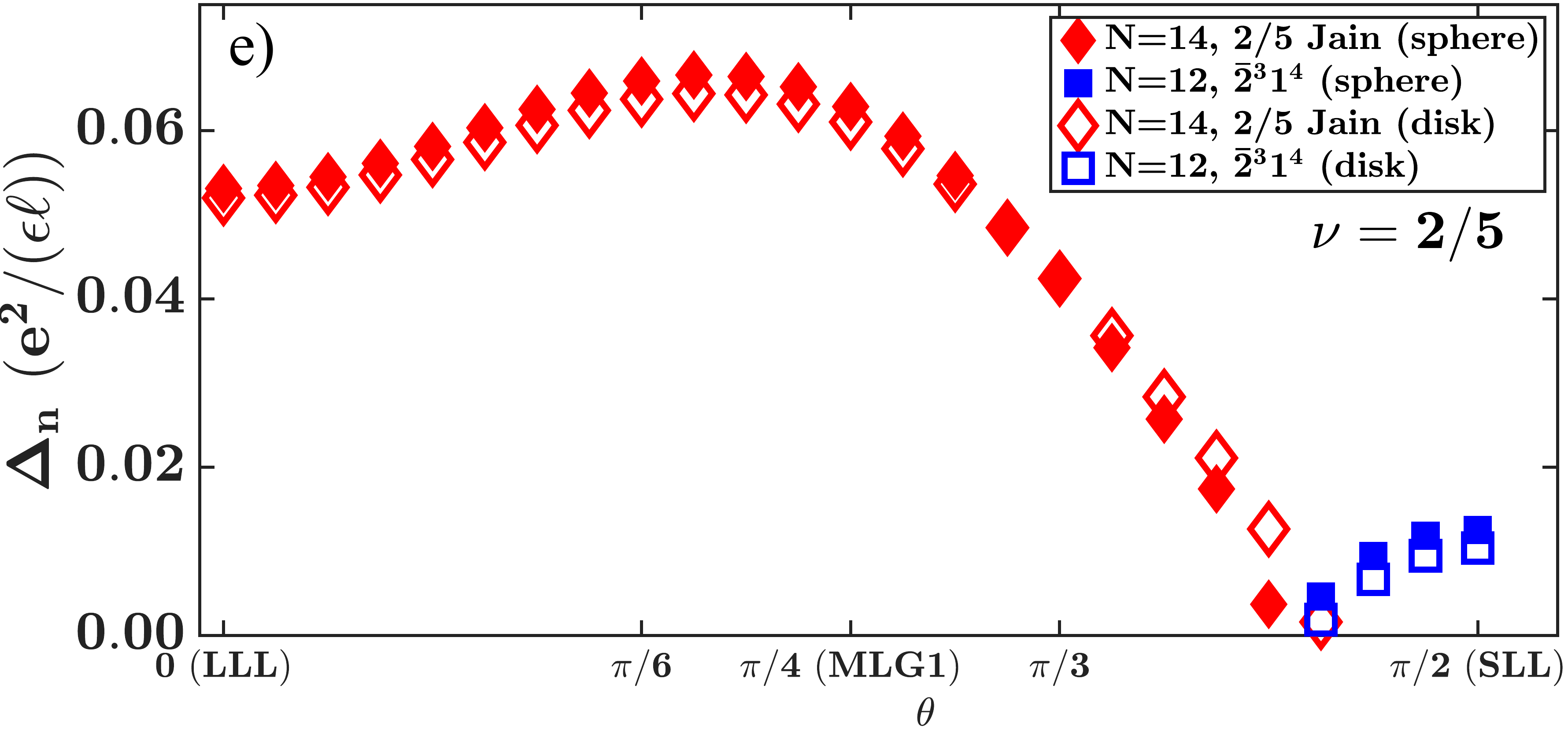}
				\includegraphics[width=0.329\textwidth]{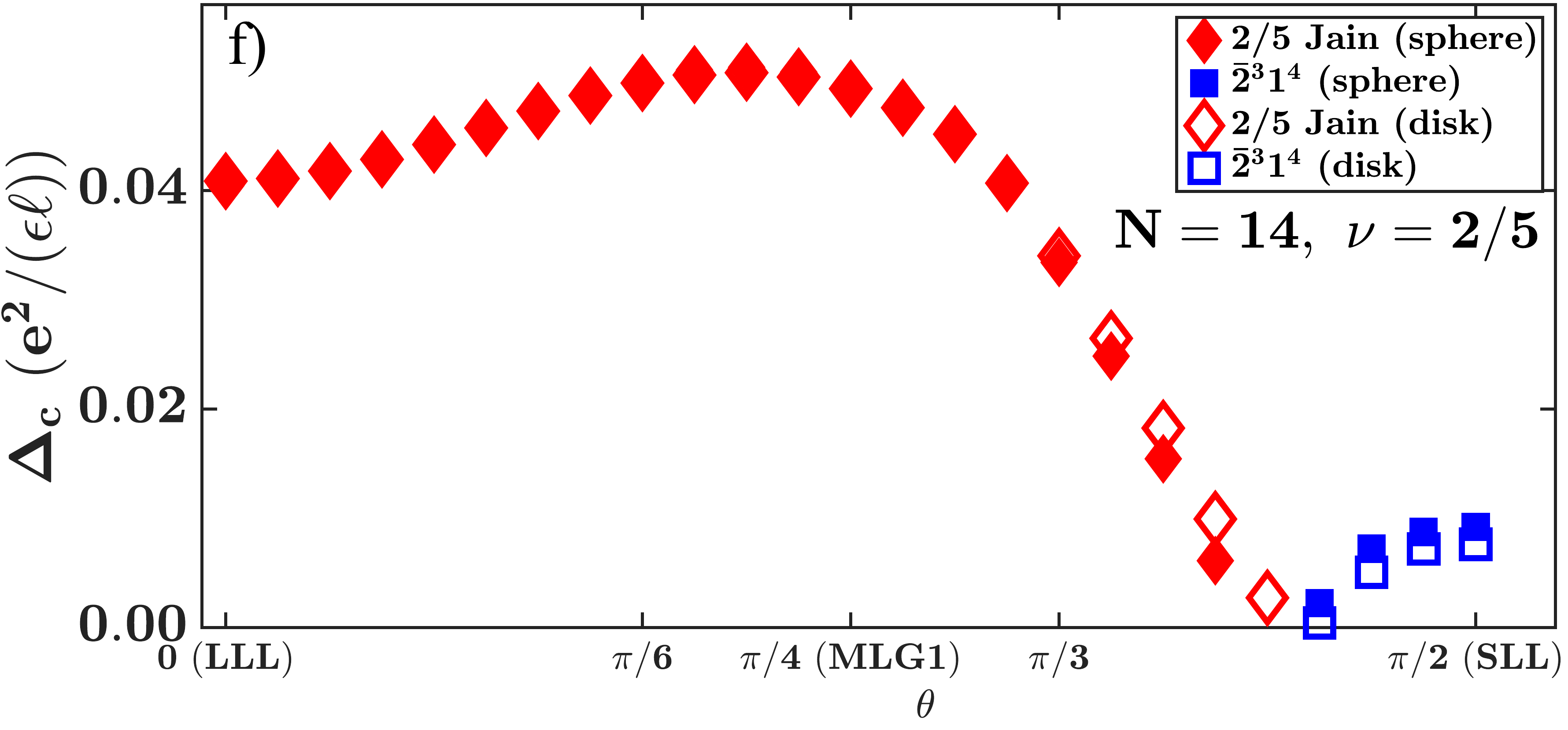} \\
				\vspace{0.3cm}
				\includegraphics[width=0.329\textwidth]{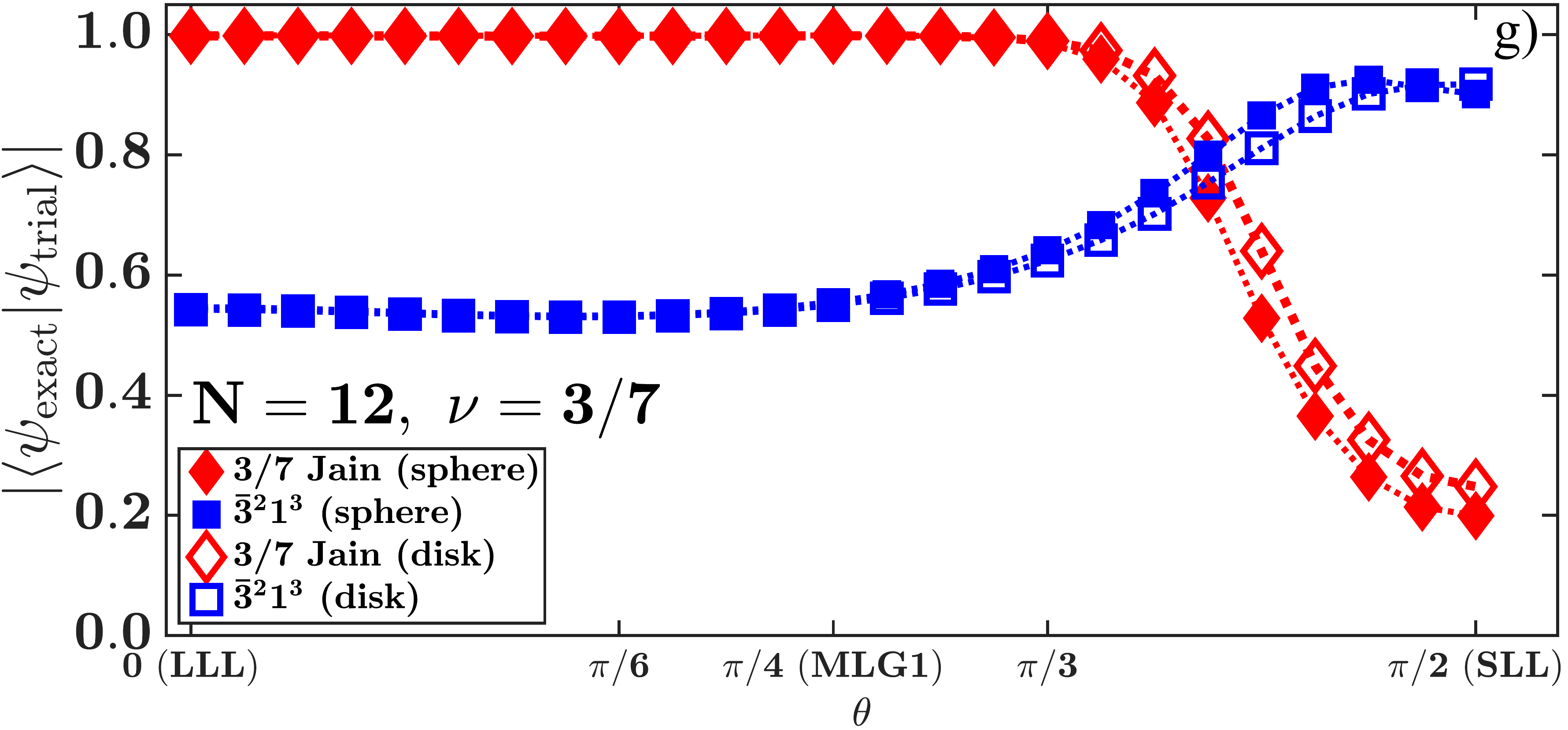}
				\includegraphics[width=0.329\textwidth]{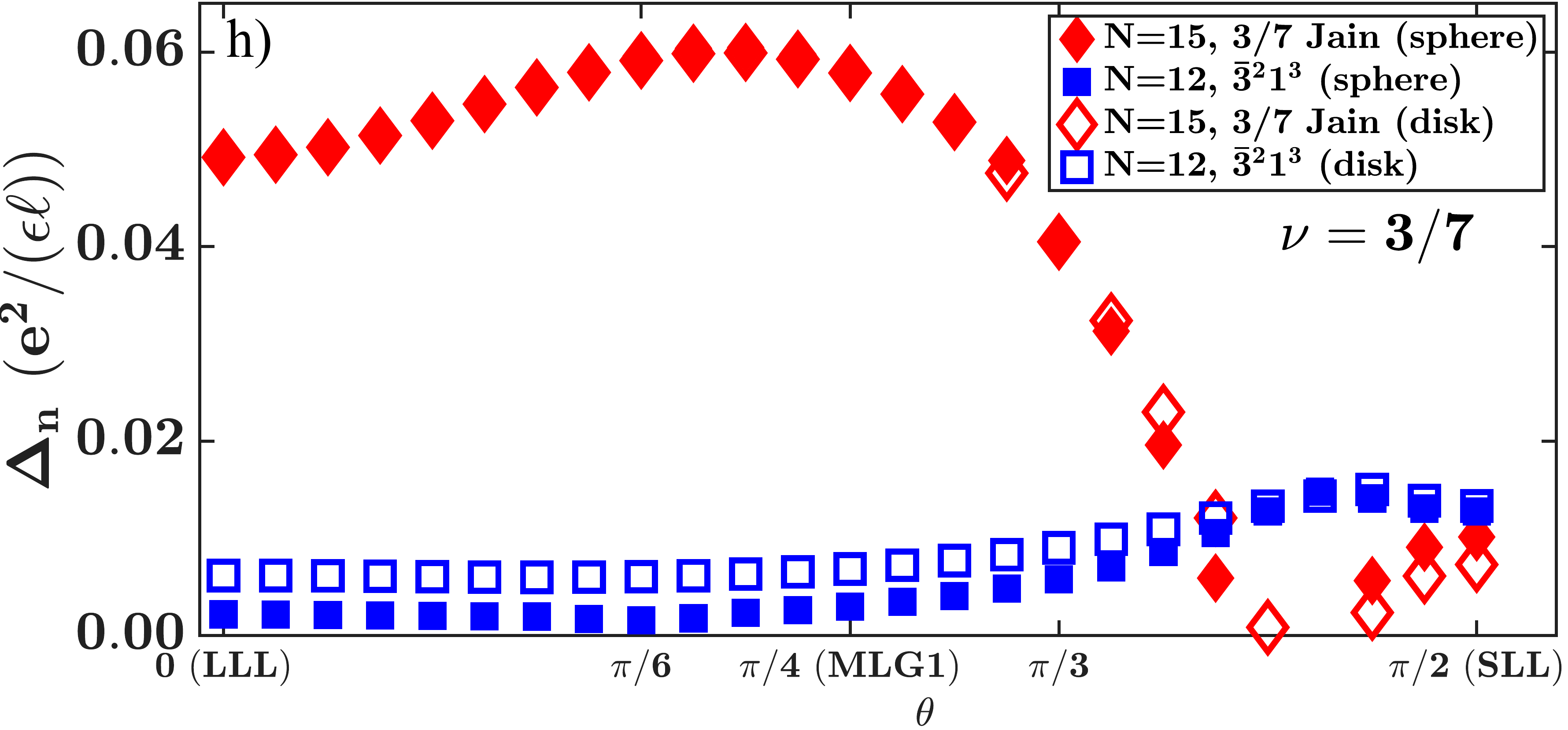}
				\includegraphics[width=0.329\textwidth]{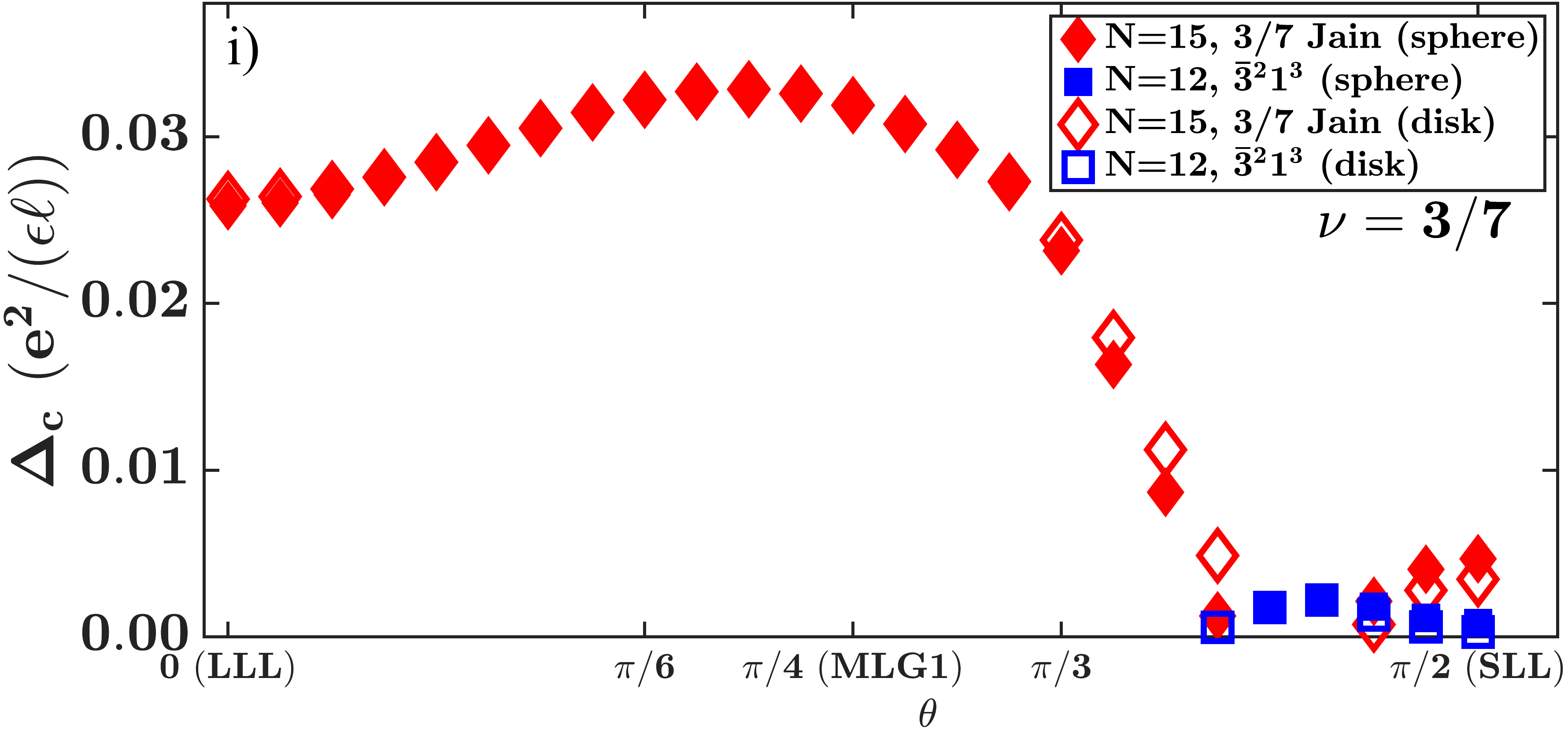}
			\end{center}
			\caption{
			Overlaps with the exact Coulomb ground state in the zeroth Landau level of bilayer graphene [(a),  (d), and (g)], neutral gaps [(b), (e), and (h)], and charge gaps [(c), (f), and (i)] as a function of the mixing angle $\theta$ that parametrized the magnetic field for candidate states at $\nu{=}1/2$ [(a)-(c)], $2/5$ [(d)-(f)] and $3/7$ [(g)-(i)] evaluated using the spherical (solid symbols) and disk (open) pseudopotentials for $N$ electrons residing on the surface of a sphere. The gaps are only shown when they are positive and the corresponding ground state is uniform.}
			\label{fig: CF_parton_transitions_ZLL_BLG}
		\end{figure*}

An important feature of an FQHE state is that it is incompressible, i.e., it has a nonzero gap to charge and neutral excitations. The charge gap, which can be accessed in transport experiments, gives the energy cost to create a far-separated pair of fundamental (smallest magnitude charge) quasiparticle and quasihole. From exact diagonalization (ED), the charge gap for a system in which the ground state of $N$ electrons occurs at shell-angular momentum $2l$ can be obtained as $\Delta_{c}{=}\left[\mathcal{E}(2l{-}1){+}\mathcal{E}(2l{+}1){-}2\mathcal{E}(2l)\right]/n_{q}$, where $\mathcal{E}(2l){=}E_{0}(2l){-}N^{2} \mathcal{C}(2l)/2$. Here $E_{0}(2l)$ is the Coulomb energy of the ground state of $N$ electrons at $2l$, $\mathcal{C}(2l)$ is the average charging energy at $2l$ which accounts for the contribution of the background~\cite{Balram20}, and $n_{q}$ is the number of fundamental quasiparticles (quasiholes) created upon the removal (insertion) of a flux quantum in the ground state. The neutral gap $\Delta_{n}{=}E_{1}(2l){-}E_{0}(2l)$ is the difference in energies of the two lowest-lying states at the $N$ and $2l$ corresponding to the ground state. All the gaps are quoted in units of $e^2/(\epsilon\ell)$, where $\epsilon$ is the dielectric constant of the host. 
We map the FQHE problem in the ZLL of BLG to a problem of electrons in the LLL interacting with a set of pseudopotentials $\{V_{m}\}$~\cite{Haldane83}, where $V_{m}$ is the energy penalty for placing two electrons in a relative angular momentum $m$ state in the ZLL of BLG. To allow for some variation in the interaction we shall carry out ED using the spherical and planar disk pseudopotentials [see Supplemental Material (SM)~\cite{SM}].

We obtain the Jain and CFFS states for small systems using a brute-force projection to the LLL. The parton states are constructed by evaluating \emph{all} the $L{=}0$ states for the corresponding system and expanding the parton wave function on the basis of all $L{=}0$ states~\cite{Balram21, Sreejith11, SM}. In Fig.~\ref{fig: CF_parton_transitions_ZLL_BLG} we show the overlaps of the exact Coulomb ground state in the ZLL of BLG with different candidate states and the charge and neutral gaps at $\nu{=}2/5,~3/7$, and $1/2$. From the high overlaps, as well as the non-zero charge gaps, we see that at low magnetic fields, i.e., in the vicinity of the SLL point, the non-Abelian parton states $\bar{2}^{3}1^{4}$, $\bar{3}^{2}1^{3}$, and $\bar{2}^{2}1^{3}$ could be stabilized. On the other hand, at higher magnetic fields the CF states prevail. The charge and neutral gaps of the parton state at $3/7$ are quite small, indicating that it is quite fragile (gaps also decrease with decreasing $B$). Strong finite-size effects are seen near the SLL point as can be deduced from the fact that we find $\Delta_{c}{<}\Delta_{n}$ while in the thermodynamic limit we expect $\Delta_{c}{\geq}\Delta_{n}$. In summary, at $\nu{=}2/5,~3/7$, and $1/2$ in the ZLL of BLG, in the vicinity of the SLL point the ground state is likely a non-Abelian parton state while in the rest of the parameter space, which includes LLL and MLG1 points, the ground state is a CF state.

\underline{\emph{Discussion.}}
In a recent experiment~\cite{Huang21} strong signatures of FQHE states at $\nu{=}1/2,~2/5$, and $3/7$ have been reported in the $\mathcal{N}{=}1$ LLs of ultra-high-quality BLG devices. Owing to the FQHE observed at half filling and a concomitant absence of it at many of the Jain fractions near $\nu{=}1/2$, we propose that these plateaus could be described by the parton states considered in this work. From Fig.~\ref{fig: CF_parton_transitions_ZLL_BLG} we estimate the critical mixing angle at which the transition from an Abelian CF state to a non-Abelian parton state occurs at all three fillings to be in the vicinity of $\theta_{c}{=}5\pi/12$. For typical parameters of graphene, this critical mixing angle corresponds to a magnetic field of $B_{c}{\approx}7$ T. Since we have considered a simplified model and made several assumptions, this value of the critical field should only be viewed as a ballpark estimate. 

Now we discuss various experimentally measurable properties that can distinguish the CF and parton states. The non-Abelian nature of the $\bar{3}^{2}1^{3}$ and $\bar{2}^{3}1^{4}$ states does not cause a further fractionalization of their quasiparticle charge. This should be contrasted with the $\bar{2}^{2}1^{3}$ state which does support a quasiparticle of charge $({-}e)/4$ at $\nu{=}1/2$. Therefore, at $\nu{=}2/5$ and $3/7$ the fundamental quasiparticles of the parton and CF states both carry the same charge of $({-}e)/5$ and $({-}e)/7$, respectively 

Due to the presence of the $\bar{2}$ and $\bar{3}$ factors the parton states are expected to host upstream edge modes which can be detected experimentally~\cite{Bid10, Dolev11, Kumar21}.  In contrast, the CF states only carry downstream edge modes. Assuming a full equilibration of the edge states, the thermal Hall conductance $\kappa_{xy}$ of an FQHE state at temperatures $T$ much smaller than the gap is expected to be quantized as $\kappa_{xy}=c_{-}[\pi^2 k^{2}_{B}/(3h)]T$, where $c_{-}$ is the chiral central charge~\cite{Kane97}.  The chiral central charge of all the CF states is integral while those of the parton states we considered fractional (see Table~\ref{tab: states_ZLL_BLG}). Recently, thermal Hall measurements have been carried out at many filling factors in GaAs~\cite{Banerjee17,  Banerjee17b} and graphene~\cite{Srivastav19, Srivastav21}. An extension of these experiments to the ZLL of BLG could help detect the partonic topological order.

\begin{table}[htpb]
\centering
\begin{tabular}{|c|c|c|c|c|c|c|}
\hline 
$B$-field & $\nu$ & state 						& nature of state & $\mathcal{S}$ & $\kappa_{xy}$ &	$\mathcal{Q}_{\rm qp}$	\\ \hline\hline
$[0, B_{c})$  & 2/5    & $\bar{2}^{3}1^{4}$  & non-Abelian &   -2				&  -4/5  		& 1/5  \\ \hline  
$(B_{c},\infty)$ & 2/5    & $211\equiv2/5$ Jain 			& Abelian	&   4					&  2  				& 1/5  \\ \hline \hline
$[0, B_{c})$ &  3/7    & $\bar{3}^{2}1^{3}$   & non-Abelian &  -3				&  -11/5			& 1/7  \\ \hline 
$(B_{c},\infty)$   &  3/7    & $311\equiv 3/7$ Jain 		& Abelian		&   5					&  3  				& 1/7  \\ \hline \hline
$[0, B_{c})$   &  1/2    & $\bar{2}^{2}1^{3}$   &  non-Abelian & -1				    &  -1/2			& 1/4  \\ \hline
$(B_{c},\infty)$  &  1/2    & FS$11$,CFFS 		& Abelian		&   2					&  -  				&	0   \\ \hline
\end{tabular} 
\caption{\label{tab: states_ZLL_BLG} The table gives some experimentally measurable properties of the various states that can arise at filling factors $\nu{=}1/2,~3/7$, and $2/5$ in the zeroth Landau level (ZLL) of bilayer graphene (BLG) as the magnetic field $B$ is varied. The states are labeled using the notation given in Eq.~\eqref{eq:parton_wf}. Using a simplified model for the interaction in the ZLL of BLG we estimate the critical value of the magnetic field $B_{c}$ at which a transition from an Abelian composite fermion (CF) to a non-Abelian parton state is $B_{c}{\approx}7$ T at all three fillings. The shift $\mathcal{S}$ on the sphere is related to the Hall viscosity $\eta_{H}{=}\hbar \nu \mathcal{S}/(8\pi \ell^2)$, $\kappa_{xy}$ is the thermal Hall conductance in units of $[\pi^2 k^{2}_{B}/(3h)]T$ (filled LLs provide an additional integral contribution) and $\mathcal{Q}_{\rm qp}$ is the charge of the fundamental (smallest charged in magnitude) quasiparticle in units of the electron charge.  The thermal Hall conductance of the CF Fermi sea (CFFS) is not expected to be quantized to a universal value since its bulk is gapless.}
\end{table}

The Hall viscosity, which measures the stress response of the FQHE state to perturbations of the underlying metric, is also anticipated to be quantized as~\cite{Read09} $\eta_{H}{=}\hbar\rho \mathcal{S}/4$, where $\rho{=}\nu/(2\pi\ell^2)$ is the electronic density and $\mathcal{S}$ is the shift~\cite{Wen92}. Since the parton and CF states have different shifts, they carry different Hall viscosities.  In Table~\ref{tab: states_ZLL_BLG} we have summarized these plausibly experimentally accessible properties of the CF and parton states at $\nu{=}1/2,~2/5$, and $3/7$. We note here that these phase transitions can potentially be studied using field theoretic techniques~\cite{Goldman20}. 

\begin{figure}[htbp]
  \centering
  \includegraphics[width=\linewidth]{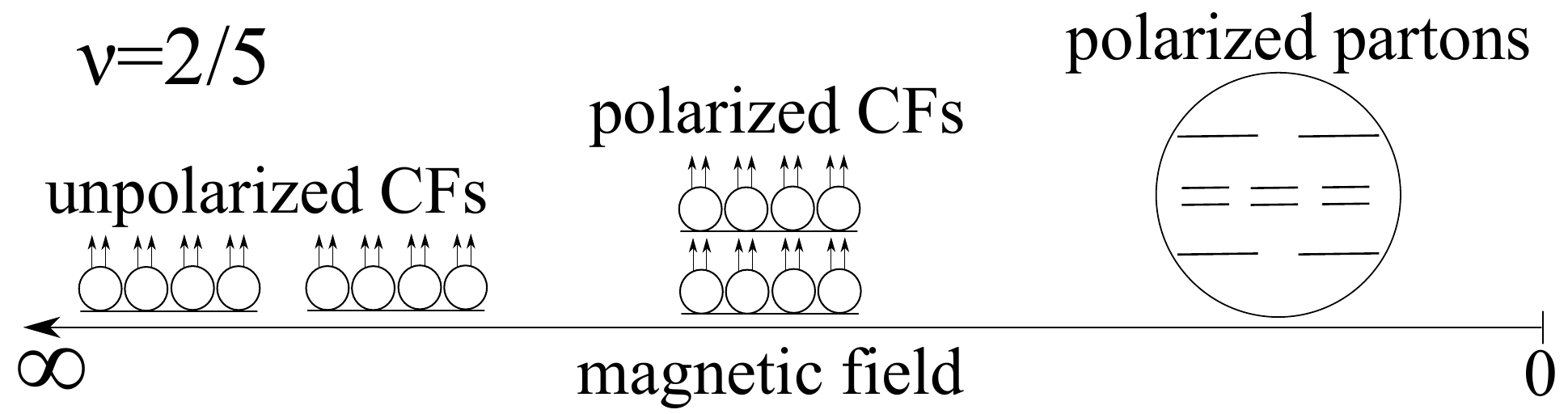}
  \caption{Schematic representation of the candidate fractional quantum Hall phases that can arise at filling $\nu{=}2/5$ in the zeroth Landau level of bilayer graphene as a function of the magnetic field.  The circle and arrows together denote a composite fermion (CF) which is a bound state of an electron (circle) and two vortices (arrows).  The partonic substructure is shown by the various partons filling different numbers of Landau levels (lines) inside the electron. In this work, we consider a single-component system and thereby focus solely on the transition between the polarized CF and parton states.}
  \label{fig: schematic_states_ZLL_BLG}
\end{figure}

We mention here that in the spinful LLL, the ground state at $\nu{=}2/5$ is a spin-singlet Jain state~\cite{Balram15a, Balram17}. On the other hand, even in the spinful $\mathcal{N}{=}1$ LL of MLG, the ground state is fully polarized, i.e., the interactions in the first excited LL of MLG are such that the CFs spontaneously polarize~\cite{Balram15c}. Therefore, in a two-component system in the ZLL of BLG, as the magnetic field is lowered, the FQHE state at $\nu{=}2/5$ transitions from a spin-singlet CF state to a fully polarized one and eventually at low magnetic fields goes to a parton state. These transitions are schematically depicted in Fig.~\ref{fig: schematic_states_ZLL_BLG}. Similarly, at $\nu{=}3/7$ in the ZLL of BLG, at large magnetic fields, the ground state would be a partially polarized Jain state. Likewise, at $\nu{=}1/2$, as the interaction is continuously tuned from the LLL to the SLL points in the half-filled ZLL of BLG, the unpolarized CFs first polarize, and then the polarized CFs pair up to form a $p$-wave superconducting state~\cite{Papic14, Zhu20a}.

We have not considered $\nu{=}1/3$ here since at all three special points, namely LLL, MLG1, and SLL, the ground state at one-third filling is believed to be Abelian~\cite{Balram13b, Jeong16, Kusmierz18, Balram20b}. Thus it is unlikely that a non-Abelian state is stabilized in the ZLL of BLG at $\nu{=}1/3$. In the SM~\cite{SM}, we have considered transitions between different \emph{Abelian} states at $\nu{=}1/3$ in the ZLL of BLG. The $s/(4s\pm 1)$ states, such as at $\nu{=}1/5$, $2/7$, and $2/9$, reside in the same topological phase as the corresponding Jain state at all three special points~\cite{Ambrumenil89, Kusmierz18, Balram21, Kim19}. Thus, we expect the topological nature of the ground state at these fillings does not change as we transition from the very high to very low magnetic field limits in the ZLL of BLG.  

In conjunction with previous works, our results show that for all the experimentally observed plateaus promising candidate parton wave functions can be constructed. Furthermore, it appears that the parton theory is sufficiently rich to capture all FQHE orders. More generally, it would be interesting to explore the possibility that structures inspired by the parton construction could aid in understanding other strongly correlated systems.

\underline{\emph{Acknowledgments.}}
We acknowledge useful discussions with William N. Faugno, Wei-Han Hsiao, Ke Huang, Jainendra K. Jain,  Zlatko Papi\'c, Arkadiusz W\'ojs, and Jun Zhu.  We acknowledge the Science and Engineering Research Board (SERB) of the Department of Science and Technology (DST) for funding support via the Start-up Grant No. SRG/2020/000154. The numerical calculations reported in this work were carried out on the Nandadevi supercomputer, which is maintained and supported by the Institute of Mathematical Science's High-Performance Computing Center. Some of the computations were performed using the DIAGHAM package, for which we are grateful to its authors.

\onecolumngrid

\setcounter{figure}{0}
\setcounter{equation}{0}
\setcounter{section}{0}
\setcounter{table}{0}
\renewcommand\thefigure{S\arabic{figure}}
\renewcommand\thetable{S\arabic{table}}
\renewcommand\thesection{S\arabic{section}}
\renewcommand\theequation{S\arabic{equation}}

\vspace{0.5cm}
\textbf{Supplemental Material for ``Transitions from Abelian composite fermion to non-Abelian parton fractional quantum Hall states in the zeroth Landau level of bilayer graphene"}
\vspace{0.5cm}

In this Supplemental Material, we present (i) details of the disk (Sec.~\ref{app: disk_pps_ZLL_BLG}) and spherical (Sec.~\ref{app: sphere_pps_ZLL_BLG}) Coulomb pseudopotentials in the zeroth Landau level (ZLL) of bilayer graphene (BLG), (ii) overlaps of the $\nu{=}1/2$ anti-Pfaffian (aPf) and $\nu{=}2/5$ anti-Read-Rezayi 3-cluster (aRR$3$) states with the exact Coulomb ground state for the largest systems accessible to us [Sec.~\ref{app: overlaps_aPf_aRR3}], and (iii) results at $\nu{=}1/3$ in the ZLL of BLG (Sec.~\ref{app: FQHE_1_3_ZLL_BLG}).

\section{Pseudopotentials in the zeroth Landau level of bilayer graphene} 
\label{app: pps_ZLL_BLG}

In this section, we present details of the disk (Sec.~\ref{app: disk_pps_ZLL_BLG}) and spherical (Sec.~\ref{app: sphere_pps_ZLL_BLG}) Coulomb pseudopotentials in the zeroth-Landau level (ZLL) of bilayer graphene (BLG). These pseudopotentials were used in the exact diagonalization (ED) calculations.

\subsection{Disk pseudopotentials}
\label{app: disk_pps_ZLL_BLG}
In this subsection, we present some formulae for the disk pseudopotentials in the ZLL of BLG. These disk pseudopotentials were used to carry out ED in the~\emph{spherical} geometry. As in the main text, the ZLL of BLG refers to the state described by the spinor wave function $(\sin(\theta)\phi_{1},\cos(\theta)\phi_{0})$, where $\phi_{n}$ is the single-particle state of non-relativistic electrons in the LL indexed by $n$ and $\theta$ is the mixing-angle which can be controlled by the external magnetic field $B$. The form-factor, which completely specifies the interaction, in the ZLL of BLG is given by:
\begin{equation}
F^{{\rm ZLL-BLG}}(\theta,q) = \Bigg[\sin^{2}(\theta)L_{1}\Big(\frac{q^2\ell^2}{2}\Big)+\cos^{2}(\theta)L_{0}\Big(\frac{q^2\ell^2}{2}\Big) \Bigg]^2,
\label{eq:ZerothLL_BLG_form_factor}
\end{equation}
where $L_{k}(x)$ is the Laguerre polynomial of degree $k$ and $\ell{=}\sqrt{\hbar c/(eB)}$ is the magnetic length. In Eq.~\eqref{eq:ZerothLL_BLG_form_factor}, it suffices to consider the range $0{\leq}\theta{\leq}\pi/2$ since the form-factor only depends on $\sin^{2}(\theta)$. For $\theta{=}0$ and $\theta{=}\pi/2$ we recover the form-factors for the lowest Landau level (LLL) and second Landau level (SLL) of GaAs respectively. The mid-point $\theta{=}\pi/4$ corresponds to the form-factor in the $\mathcal{N}{=}1$ LL of monolayer graphene (MLG1)~\cite{Balram15c}. The spherically symmetric $1/r$ Coulomb interaction between electrons in any LL is conveniently parametrized using the Haldane pseudopotentials~\cite{Haldane83} $V_m$, which is the energy of two electrons in a state of relative angular momentum $m$ (for single-component systems like the one we consider, only the odd $m$ pseudopotentials are relevant). The Haldane pseudopotentials in the ZLL of BLG as a function of $\theta$ in the planar disk geometry are given by:
\begin{equation}
V^{{\rm ZLL-BLG}}_{m}(\theta){=}\int\frac{d^{2}\vec{q}}{(2\pi)^2}\frac{2\pi e^{2}}{\epsilon q}F^{{\rm ZLL-BLG}}(\theta,q) e^{-q^2 \ell^2} L_{m}(q^2 \ell^2) {=}\frac{e^{2}}{\epsilon}\int_{0}^{\infty} dq\Bigg[\sin^{2}(\theta)L_{1}\Big(\frac{q^2\ell^2}{2}\Big)+\cos^{2}(\theta)L_{0}\Big(\frac{q^2\ell^2}{2}\Big) \Bigg]^2 e^{-q^2\ell^2} L_{m}(q^2\ell^2)
\label{eq:Haldane_pps_ZerothLL_BLG}
\end{equation}
where $q{=}|\vec{q}|$ is the magnitude of the planar wave vector $\vec{q}$ and $2\pi e^{2}/(\epsilon q)$ is the Fourier transform of the Coulomb interaction with $\epsilon$ the dielectric constant of the background host material. The integral in Eq.~\eqref{eq:Haldane_pps_ZerothLL_BLG} can be evaluated analytically and results in the following expression for the Coulomb pseudopotentials in the disk geometry for the ZLL of BLG:
\begin{equation}
V^{{\rm ZLL-BLG}}_{m}(\theta)=\frac{\sqrt{\pi }}{32} \Bigg(16~_2F_1\left(\frac{1}{2},-m;1;1\right)-8~_2F_1\left(\frac{3}{2},-m;1;1\right)\sin^{2}(\theta)+3~_2F_1\left(\frac{5}{2},-m;1;1\right)\sin^{4}(\theta) \Bigg)~\frac{e^{2}}{\epsilon\ell},
\label{eq:ZerothLL_BLG_disk_pps}
\end{equation}
where $_2F_1$ is the Gauss hypergeometric function. The charging energy, which is used to evaluate the contribution of the positively charged background (see supplemental material of Ref.~\cite{Balram20b}), corresponding to the above disk pseudopotentials is given by:
\begin{equation}
\mathcal{C}^{(0,|1\rangle)}(2l,\theta)=\frac{\Gamma \left((2l)-\frac{1}{2}\right) (12 (8l (32l-7)-5) \cos (2 \theta )+9 (8l-1) \cos (4 \theta )+8l (128l (16l+3)-53)-315)}{1536 (2l+1) \Gamma (2l+2)}~\frac{e^{2}}{\epsilon\ell},
\label{eq:ZerothLL_BLG_disk_pps_charging_energy}
\end{equation}
where $\Gamma(x)$ is the Gamma function, $l{=}|Q|+\mathcal{N}$ is the shell angular momentum (the number of orbitals in the spherical geometry is $2l{+}1$), $2Q$ is the flux through the sphere and $\mathcal{N}$ is the LL of our interest ($\mathcal{N}=1$ in the ZLL of BLG). The charging energy is required to estimate the charge gap of a state. In the spherical geometry, a truncated set of these disk pseudopotentials with $0{\leq}m{\leq}2l$, are used to carry out ED.

\subsection{Spherical pseudopotentials}
\label{app: sphere_pps_ZLL_BLG}
In this subsection, we present some details on the evaluation of spherical pseudopotentials in the ZLL of BLG. The spinor matrix element of an interaction $V$ in the spherical geometry in the LL described by the form-factor of Eq.~\eqref{eq:ZerothLL_BLG_form_factor} can be expressed in terms of the usual matrix elements of non-relativistic LLs as follows:
\begin{eqnarray}
\left(1',2'|V|1,2 \right) \equiv \left<\left<m_1,m_2||V||m_3,m_4\right>\right>
=\left[\cos^{2}(\theta)~\cos^{2}(\theta)\right]&\left<0,m_1;0,m_2|V|0,m_3;0,m_4\right>& \\
+\left[\cos^{2}(\theta)~\sin^{2}(\theta)\right]&\left<0,m_1;1,m_2|V|0,m_3;1,m_4\right>& \nonumber\\
+\left[\sin^{2}(\theta)~\cos^{2}(\theta)\right]&\left<1,m_1;0,m_2|V|1,m_3;0,m_4\right>& \nonumber\\
+\left[\sin^{2}(\theta)~\sin^{2}(\theta)\right]&\left<1,m_1;1,m_2|V|1,m_3;1,m_4\right>&.\nonumber
\label{eq:spinor_matrix element_ZerothLL_BLG_sphere}
\end{eqnarray}
All the orbitals appearing in the above matrix elements have the same $l$, which implies the flux value is variable with $Q{=}l{-}1$ or $l$. The pseudopotentials of a spherically symmetric interaction $V(r)$ are given by
\begin{equation}
 V_{L}=\sum_{m_{1}=-l}^{l}\sum_{m_{2}=-l}^{l}\sum_{m'_{1}=-l}^{l}\sum_{m'_{2}=-l}^{l} \langle L,m|l,m'_{1};l,m'_{2}\rangle \langle l,m_{1};l,m_{2}|L,m\rangle ~\left(1',2'|V(r)|1,2 \right)~\delta_{m,m_{1}+m_{2}}\delta_{m_{1}+m_{2},m'_{1}+m'_{2}},
\end{equation}
where $l{=}|Q|{+}1$ is the shell-angular momentum and $\langle j_{1},m_{1};j_{2},m_{2}|j_{3},m_{3}\rangle$ is the Clebsch-Gordan coefficient. The total orbital angular momentum $L$ is related to the relative orbital angular momentum $m$ as $m{=}2l-L$. For the Coulomb interaction, $V(r){=}1/r$, the spinor matrix element is given by
\begin{eqnarray}
\left(1',2'|\frac{1}{r}|1,2 \right) 
=\left[\cos^{2}(\theta)~\cos^{2}(\theta)\right]~&V_{C}(Q+1,Q+1,m_{1},m_{2},m'_{1},m'_{2},Q,1)& \\
+\left[\cos^{2}(\theta)~\sin^{2}(\theta)\right]~&V_{C}(Q+1,Q+0,m_{1},m_{2},m'_{1},m'_{2},Q,1)& \nonumber \\
+\left[\sin^{2}(\theta)~\cos^{2}(\theta)\right]~&V_{C}(Q+0,Q+1,m_{1},m_{2},m'_{1},m'_{2},Q,1)& \nonumber \\
+\left[\sin^{2}(\theta)~\sin^{2}(\theta)\right]~&V_{C}(Q+0,Q+0,m_{1},m_{2},m'_{1},m'_{2},Q,1)&,\nonumber
\label{eq:spinor_matrix element_ZerothLL_BLG_sphere_Q}
\end{eqnarray}
where $V_{C}(Q_{1},Q_{2},m_{1},m_{2},m'_{1},m'_{2},Q,1)$ is the two-body Coulomb matrix element for a pair of non-relativistic fermions, which is given by
\begin{eqnarray}
V_{C}(Q_{1},Q_{2},m_{1},m_{2},m'_{1},m'_{2},Q,1)&=&\frac{e^{2}}{\epsilon R}(2l+1)^{2}(-1)^{Q_1+Q_2-m_1'-m_2'} \times \\
&&\sum_{l'=0}^{2l}\sum_{m'=-l'}^{l'}(-1)^{m'}\begin{pmatrix} l & l' & l \\ m_1' & m' & -m_1\end{pmatrix}\begin{pmatrix} l & l' & l \\ -Q_1 & 0 & Q_1\end{pmatrix}\begin{pmatrix} l & l' & l \\ m_2' & -m' & -m_2\end{pmatrix}\begin{pmatrix} l & l' & l \\ -Q_2 & 0 & Q_2\end{pmatrix}, \nonumber
\end{eqnarray}
where $\left([j_{1},j_{2},j_{3}];[m_{1},m_{2},m_{3}]\right)$ is the Wigner $3j$ symbol and $R{=}\ell\sqrt{l}$ is our choice for the radius of the sphere~\cite{Balram20b} with $\ell{=}\sqrt{\hbar c/(eB)}$ being the magnetic length. For a fast and accurate evaluation of the Clebsch-Gordan coefficients and the Wigner $3j$ symbols we use the algorithm outlined in Ref.~\cite{Johansson16}.

\section{Results for larger systems} 
\label{app: overlaps_aPf_aRR3}
In the main text, we showed results for the $\bar{2}^{2}1^{3}$ state, described by the wave function $\mathcal{P}_{\rm LLL}[\Phi^{*}_{2}]^{2}\Phi^{3}_{1}{\sim}[\Psi^{\rm Jain}_{2/3}]^{2}/\Phi_{1}$ (The ${\sim}$ indicates that states either side of the sign differ in the details of how the projection to the LLL is implemented. We expect such details to not alter the topological phase of the underlying state~\cite{Balram16b}.), for $N{=}14$ electrons. In this subsection, we present some details on the construction of the wave function that was used in obtaining those results. To construct the state in Fock-space, we follow the method outlined in Refs.~\cite{Balram20a, Sreejith11}, which involves the calculation of all the uniform states on the sphere, i.e., with total orbital angular momentum $L{=}0$, for the corresponding system and then obtaining the projection of the parton state on these uniform states. We have not been able to construct the constituent $\nu=2/3$ Jain state in the Fock-space for $N{=}14$ electrons. However, we can obtain a very good approximation to the 2/3 Jain state by using either the anti-Laughlin 2/3 state, $\Psi^{\rm a-Laughlin}_{2/3}$, or just the LLL Coulomb ground state at 2/3, $\Psi^{0{\rm LL}}_{2/3}$. The anti-Laughlin 2/3 state, which is the particle-hole conjugate of the 1/3 Laughlin and the 2/3 LLL Coulomb ground states has near unit overlap with the 2/3 Jain state for all systems where the overlap calculation can be carried out (see Table~\ref{tab:overlaps_n_0_LL_bar211_aLaughlin}). Unlike the 2/3 Jain state, which has to be obtained by a brute-force projection to the LLL (which is doable for up to $N{=}12$ electrons), the anti-Laughlin 2/3 and LLL Coulomb ground states can be constructed by brute-force ED (which can be done for up to $N{=}28$ electrons).  In the main text, the results shown for the $\bar{2}^{2}1^{3}$ state for $N{=}14$ electrons were obtained by constructing the wave function $[\Psi^{\rm a-Laughlin}_{2/3}]^{2}/\Phi_{1}$.

\begin{table}[h]
\centering
\begin{tabular}{|c|c|c|c|c|}
\hline
$N$  & $2l$ & $|\langle\Psi^{\rm Jain}_{2/3}|\Psi^{\rm a-Laughlin}_{2/3} \rangle|^{2}$ & $|\langle \Psi^{0{\rm LL}}_{2/3}|\Psi^{{\rm Jain}}_{2/3} \rangle|^{2}$  &  $|\langle \Psi^{0{\rm LL}}_{2/3}|\Psi^{\rm a-Laughlin}_{2/3} \rangle|^{2}$ \\ \hline
4	&	6	&	1.000	&	1.000	&	1.000	\\ \hline
6	&	9	&	1.000	&	0.993	&	0.996	\\ \hline
8	&	12	&	1.000	&	0.996	&	0.998	\\ \hline
10	&	15	&	0.999	&	0.988	&	0.993	\\ \hline
12	&	18	&	0.999	&	0.987	&	0.993	\\ \hline
14	&	21	&	-	    &	-	    &	0.991	\\ \hline
16	&	24	&	-	    &	-	    &	0.988	\\ \hline
18	&	27	&	-	    &	-	    &	0.986	\\ \hline
20	&	30	&	-	    &	-	    &	0.984	\\ \hline
22	&	33	&	-	    &	-	    &	0.982	\\ \hline
24	&	36	&	-	    &	-	    &	0.980	\\ \hline
26	&	39	&	-	    &	-	    &	0.978	\\ \hline
28	&	42	&	-	    &	-	    &	0.975	\\ \hline
\end{tabular}
\caption{\label{tab:overlaps_n_0_LL_bar211_aLaughlin} 
Squared overlaps of the 2/3 Jain state (equivalently the $\bar{2}1^{2}$ state), $\Psi^{{\rm Jain}}_{2/3}$, with the anti-Laughlin $2/3$ state, $\Psi^{\rm a-Laughlin}_{2/3}$, for $N$ electrons in the spherical geometry. For comparison we also show squared overlaps of the Coulomb ground state in the lowest Landau level of non-relativistic electrons (obtained by exact diagonalization), $\Psi^{0{\rm LL}}_{2/3}$, with these two states. The numbers in the last column are taken from Ref.~\cite{Balram20}. The - indicates a system for which results are not available.}
\end{table}

Along the same lines, can we construct the wave function for the $\bar{2}^{2}1^{3}$ state in Fock-space for the next system of $N=16$ electrons using the 16-particle $\Psi^{\rm a-Laughlin}_{2/3}$ as the constituent state? Unfortunately, not. Our method for constructing the parton state relies on obtaining \emph{all} the uniform states. We have not been to obtain all the $L{=}0$ states for the $N{=}16$ system since its Hilbert space is quite large (dim$_{L_{z}=0}{=}29,927,526$ and dim$_{L=0}{=}17,344$). 

\begin{table}[h]
\centering
\begin{tabular}{|c|c|c|c|c|}
\hline
$N$  & $2l$ & $|\langle\Psi^{\bar{2}^{2}1^{3}}_{1/2}|\Psi^{\rm aPf}_{1/2} \rangle|^{2}$ &  $|\langle \Psi^{1{\rm LL}}_{1/2}|\Psi^{\bar{2}^{2}1^{3}}_{1/2} \rangle|^{2}$ &  $|\langle \Psi^{1{\rm LL}}_{1/2}|\Psi^{\rm aPf}_{1/2} \rangle|^{2}$ \\ \hline
4	&	9	&	0.929	&	0.885	&	0.666	\\ \hline
6	&	13	&	0.940	&	0.862	&	0.752	\\ \hline
8	&	17	&	0.908	&	0.862	&	0.702	\\ \hline
10	&	21	&	0.881	&	0.774	&	0.671	\\ \hline
12	&	25	&	0.861	&	0.614	&	0.481	\\ \hline
14	&	29	&	-	    &	-	    &	0.608	\\ \hline
16	&	33	&	-	    &	-	    &	0.458	\\ \hline
18	&	37	&	-	    &	-	    &	0.454	\\ \hline
\end{tabular}
\caption{\label{tab:overlaps_n_1_LL_bar2bar2111_aPf} 
Squared overlaps of the $1/2$ anti-Pfaffian state, $\Psi^{\rm aPf}_{1/2}$, with the $\bar{2}^{2}1^{3}$ state, $\Psi^{\bar{2}^{2}1^{3}}_{1/2}$, for $N$ electrons in the spherical geometry. For comparison we also show squared overlaps of the Coulomb ground state in the second Landau level of non-relativistic electrons (obtained by exact diagonalization), $\Psi^{1{\rm LL}}_{1/2}$, with these two states. The numbers in the last column are taken from Ref.~\cite{Balram20}. The - indicates a system for which results are not available.}
\end{table}

\begin{table}[h]
\centering
\begin{tabular}{|c|c|c|c|c|}
\hline
$N$  & $2l$ & $|\langle\Psi^{\bar{2}^{3}1^{4}}_{2/5}|\Psi^{\rm aRR3}_{2/5} \rangle|^{2}$  & $|\langle \Psi^{1{\rm LL}}_{1/2}|\Psi^{\bar{2}^{3}1^{4}}_{2/5} \rangle|^{2}$ &   $|\langle \Psi^{1{\rm LL}}_{2/5}|\Psi^{\rm aRR3}_{2/5} \rangle|^{2}$ \\ \hline
4	&	12	&	0.841	&	0.699	&	0.971	\\ \hline
6	&	17	&	0.829	&	0.462	&	0.814	\\ \hline
8	&	22	&	0.778	&	0.681	&	0.968	\\ \hline
10	&	27	&	0.704	&	0.479	&	0.878	\\ \hline
12	&	32	&	0.671	&	0.450	&	0.808	\\ \hline
14	&	37	&	-	    &	-	    &	0.656	\\ \hline
\end{tabular}
\caption{\label{tab:overlaps_n_1_LL_bar2bar2bar21111_aRR3} 
Squared overlaps of the $2/5$ anti-Read-Rezayi 3-cluster (aRR$3$) state, $\Psi^{\rm aRR3}_{2/5}$, with the $\bar{2}^{3}1^{4}$ state, $\Psi^{\bar{2}^{3}1^{4}}_{2/5}$, for $N$ electrons in the spherical geometry. For comparison we also show squared overlaps of the Coulomb ground state in the second Landau level of non-relativistic electrons (obtained by exact diagonalization), $\Psi^{1{\rm LL}}_{2/5}$, with these two states. The - indicates a system for which results are not available. }
\end{table}

Nevertheless, for the $N{=}16$ system, which is the largest one for which we can do ED, we can compare the ED results against the anti-Pfaffian (aPf) wave function since the aPf state has a good overlap with the $\bar{2}^{2}1^{3}$ state for all accessible systems~\cite{Balram18} (see Table~\ref{tab:overlaps_n_1_LL_bar2bar2111_aPf}). Similarly, for the $\bar{2}^{3}1^{4}$ state, we showed results for $N{=}12$ in the main text. For $N{=}14$, which is the largest system we can access using ED, we can compare the exact results against the anti-Read-Rezayi 3-cluster (aRR3) wave function since the aRR3 state has a good overlap with the $\bar{2}^{3}1^{4}$ state for all accessible systems~\cite{Balram19} (see Table~\ref{tab:overlaps_n_1_LL_bar2bar2bar21111_aRR3}). In this subsection, we present overlaps of the aPf and the aRR$3$ wave functions with the exact Coulomb ground states in the ZLL of BLG for these larger systems.

\begin{figure}[htpb]
			\begin{center}			
				\includegraphics[width=0.49\textwidth]{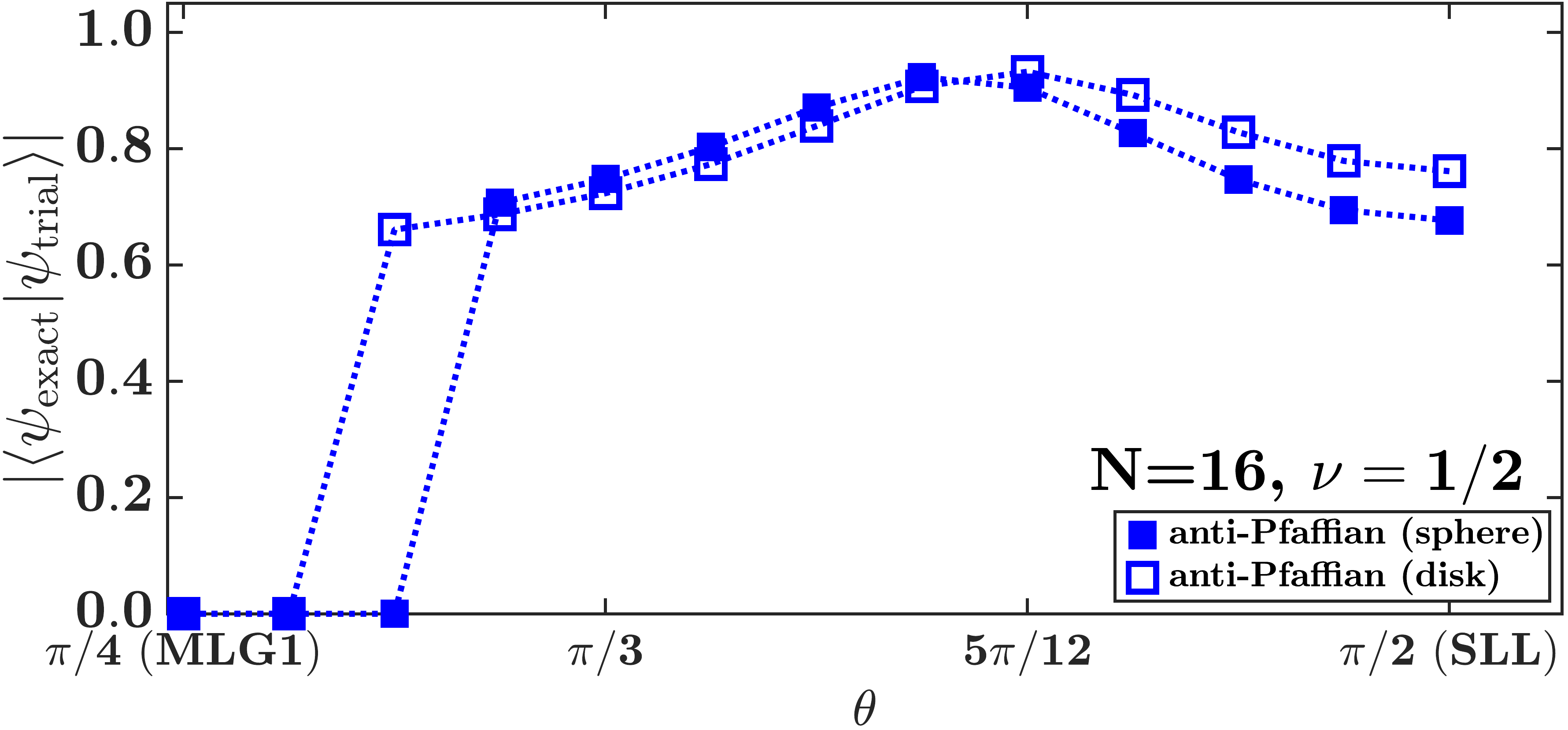}
				\includegraphics[width=0.49\textwidth]{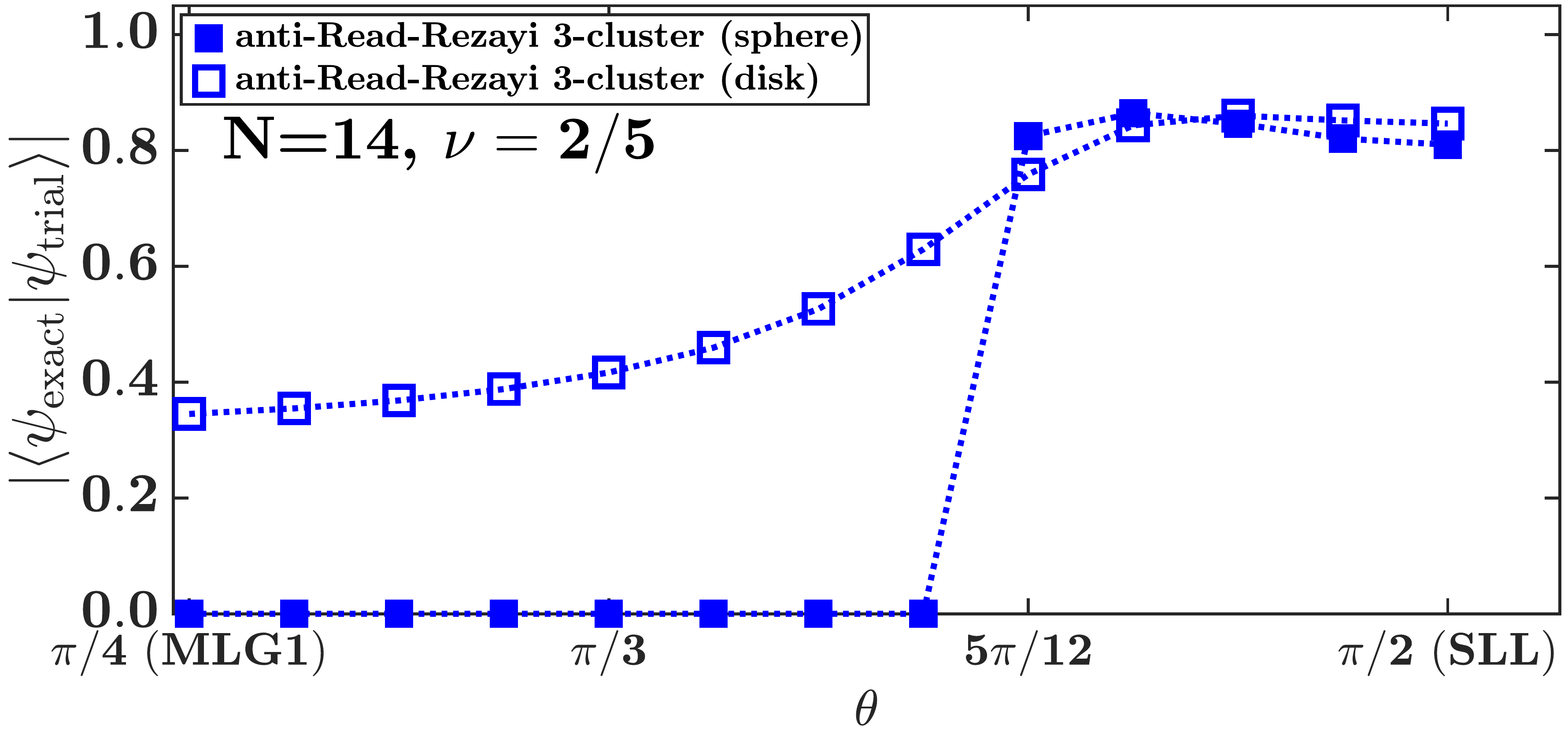}
			\end{center}
			\caption{
			Overlaps of the exact Coulomb ground state in the zeroth Landau level of bilayer graphene as a function of the mixing angle $\theta$ with the $16$-particle anti-Pfaffian state at $\nu{=}1/2$ [left panel] and the $14$-particle anti-Read-Rezayi 3-cluster state $\nu{=}2/5$ [right panel] evaluated using the spherical (filled symbols) and disk (empty) pseudopotentials in the spherical geometry.}
			\label{fig: aPf_aRR3_ZLL_BLG}
		\end{figure}

In Fig.~\ref{fig: aPf_aRR3_ZLL_BLG} we show overlaps of the exact Coulomb ground state in the ZLL of BLG with the $\nu{=}1/2$ aPf and the $\nu{=}2/5$ aRR$3$ states for $N{=}16$ and $N{=}14$ electrons respectively.  We have only calculated the overlaps between the MLG1 and SLL points since these non-Abelian states are only expected to be stabilized in this region of the parameter space. The overlaps are low close to the MLG1 point, peak near the vicinity of the predicted transition $\theta_{c}{=}5\pi/12$, and remain sizable up to the SLL point.  These results parallel the results shown in Fig.  1 of the main text where the overlaps were computed with the $\bar{2}^{2}1^{3}$ and $\bar{2}^{3}1^{4}$ states respectively. Thus, these results further corroborate the result stated in the main text that for $\theta{\gtrsim}\theta_{c}{=}5\pi/12$, the $\bar{2}^{2}1^{3}$ and $\bar{2}^{3}1^{4}$ states could be stabilized in the ZLL of BLG.
		
\section{Fractional quantum Hall effect at $\nu{=}1/3$ in the zeroth-Landau level of bilayer graphene}	
\label{app: FQHE_1_3_ZLL_BLG}
In the main text, we mentioned that we do not expect a non-Abelian state to be stabilized at $\nu{=}1/3$ in the ZLL of BLG.  A potential non-Abelian candidate state that could arise at $1/3$ is the $\bar{2}^{4}1^{5}$ state which lies in the same topological phase as the particle-hole conjugate of the $4$-cluster Read-Rezayi state~\cite{Read99, Peterson15, Balram19}.  At the shift $\mathcal{S}{=}-3$ corresponding to the $\bar{2}^{4}1^{5}$ state, the SLL Coulomb ground state is not consistently uniform.  Moreover, the thermal Hall measurements at $7/3$ of GaAs~\cite{Banerjee17b} are inconsistent with the value predicted for the $\bar{2}^{4}1^{5}$ state. Since the $\bar{2}^{4}1^{5}$ state likely does not occur in the SLL, it is also unlikely to be stabilized in the ZLL of BLG, and thus we have not considered it further.  However, there could be transitions between different \emph{Abelian} states at $\nu{=}1/3$ as the interaction is tuned from the MLG1 to the SLL point. In this section, we will discuss these transitions. 

A family of ``$\mathbb{Z}_{n}$" topologically ordered Abelian parton states, denoted by $n\bar{n}1^{3}$, and described by the wave function $[\Psi^{\rm Jain}_{n/(2n+1)}{\times}\Psi^{\rm Jain}_{n/(2n-1)}]/\Phi_{1}$ was proposed as a candidate to describe the $7/3$ fractional quantum Hall effect (FQHE), i.e., the $1/3$ FQHE in the SLL~\cite{Balram20}. The $1\bar{1}1^{3}$ state is topologically equivalent to the $\nu{=}1/3$ Laughlin state~\cite{Laughlin83}. In Ref.~\cite{Balram20}, the wave functions of the $2\bar{2}1^{3}$ and $3\bar{3}1^{3}$ states were constructed in Fock space for up to $N=10$ and $N=9$ electrons respectively. Using the method outlined in Ref.~\cite{Balram20a} (briefly reviewed in Sec~\ref{app: overlaps_aPf_aRR3}), we have now been able to obtain the Fock space representation of the $2\bar{2}1^{3}$ and $3\bar{3}1^{3}$ states, evaluated respectively as $[\Psi_{2/5}^{\rm Jain}\Psi_{2/3}^{\rm Jain}]/\Phi_{1}$ and $[\Psi_{3/7}^{\rm Jain}\Psi_{3/5}^{\rm Jain}]/\Phi_{1}$, for the system of $N{=}12$ electrons at $2l=33$. 

\begin{figure}[htpb]
			\begin{center}			
				\includegraphics[width=0.49\textwidth]{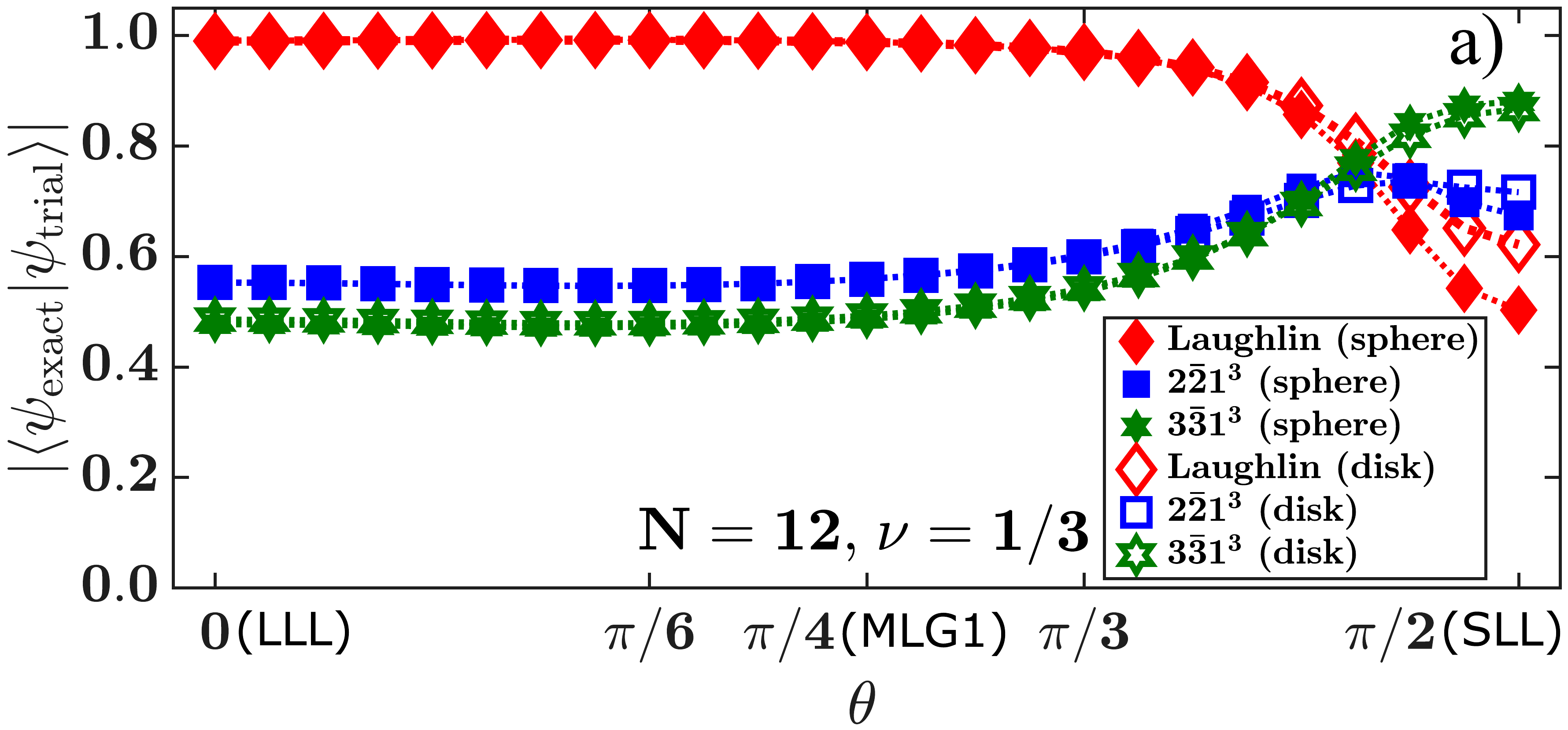}
				\includegraphics[width=0.49\textwidth]{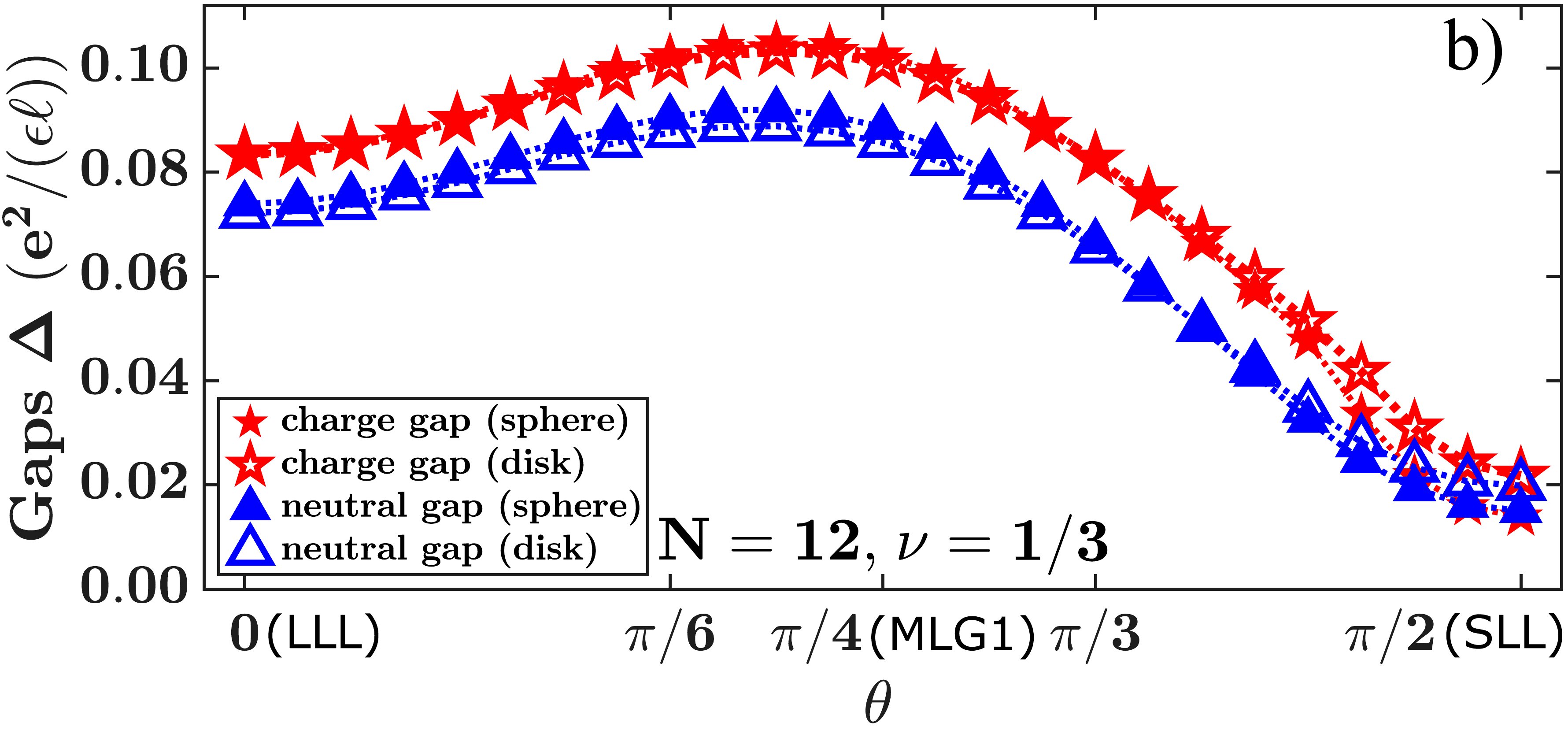}
			\end{center}
			\caption{
			Overlaps [left panel a)] of the $3\bar{3}1^{3}$ (green hexagrams), $2\bar{2}1^{3}$ (blue squares) and Laughlin (red diamond) states at $\nu{=}1/3$ with the exact Coulomb ground state and charge (red pentagrams) and neutral (blue triangles) gaps  [right panel b)] in the zeroth-Landau level of bilayer graphene as a function of the mixing angle $\theta$ evaluated in the spherical geometry using the spherical (filled symbols) and disk (open symbols) pseudopotentials for $N{=}12$ electrons. }
			\label{fig: overlaps_gaps_1_3_ZLL_BLG_3bar3111_2bar2111_Laughlin}
		\end{figure}

The authors of Ref.~\cite{Faugno20b} proposed that a transition between a $\mathbb{Z}_{n{\geq}2}$ and the Laughlin state can be observed in the ZLL of BLG as the magnetic field is lowered. By decreasing the magnetic field, the mixing angle $\theta$ can be increased, thereby the interaction can be varied from the MLG1 ($\theta{=}\pi/4$) to the SLL ($\theta{=}\pi/2$) point.  In Fig.~\ref{fig: overlaps_gaps_1_3_ZLL_BLG_3bar3111_2bar2111_Laughlin}a) we show the overlap of the exact ZLL Coulomb ground state in BLG obtained using the spherical and disk pseudopotentials with the $3\bar{3}1^{3}$, $2\bar{2}1^{3}$ and Laughlin states for a system of $N{=}12$ electrons at $2l{=}33$ as a function of the mixing angle $\theta{\in}[0,\pi/2]$.  We find that the $\mathbb{Z}_{2}$ and $\mathbb{Z}_{3}$ states have a higher overlap than the Laughlin state for $\theta{>}7\pi/16$ while the Laughlin state has a higher overlap than the $\mathbb{Z}_{n{\geq}2}$ states for $\theta{<}7\pi/16$.  This result is in agreement with the assertion made in Ref.~\cite{Faugno20b} that in the ZLL of BLG the Laughlin state at high magnetic fields (small $\theta$) could transition into a $\mathbb{Z}_{n{\geq}2}$ state at small magnetic fields (large $\theta$). Ref.~\cite{Faugno20b} estimated that this transition occurs at $\theta_{c} \sim 1.45$ and this value is quite close to $\theta{=}7\pi/16{=}1.37$ in the vicinity of which for the system of $N{=}12$ electrons the overlap of the $\mathbb{Z}_{2}$ and $\mathbb{Z}_{3}$ states pick up and the Laughlin state drops as the mixing angle is increased, i.e., the magnetic field is reduced [see Fig.~\ref{fig: overlaps_gaps_1_3_ZLL_BLG_3bar3111_2bar2111_Laughlin}a)]. Interestingly, this critical mixing angle $\theta'_{c}=7\pi/16$ is also quite close to the critical mixing angle $\theta_{c}{=}5\pi/12$ at which we predict a transition from Abelian CF states to non-Abelian parton states at nearby fillings such as $\nu{=}2/5,~3/7$ and $1/2$ (see main text). In Fig.~\ref{fig: overlaps_gaps_1_3_ZLL_BLG_3bar3111_2bar2111_Laughlin}b) we show the neutral and charge gaps (charge gap is computed assuming that the insertion of a single flux quantum creates a single quasihole) for the $1/3$ state in the ZLL of BLG. The charge gap of the $\mathbb{Z}_{n}$ state is reduced by a factor of $n$ compared to the values shown on the plot since its fundamental (smallest magnitude charge) quasiparticle has a charge $(-e)/(3n)$. The positive value of the charge gap at all values of the mixing angle suggests that FQHE at $\nu{=}1/3$ is robust at all magnetic fields in the ZLL of BLG.

\begin{figure}[htpb]
\begin{center}
\includegraphics[width=0.6\textwidth,height=0.3\textwidth]{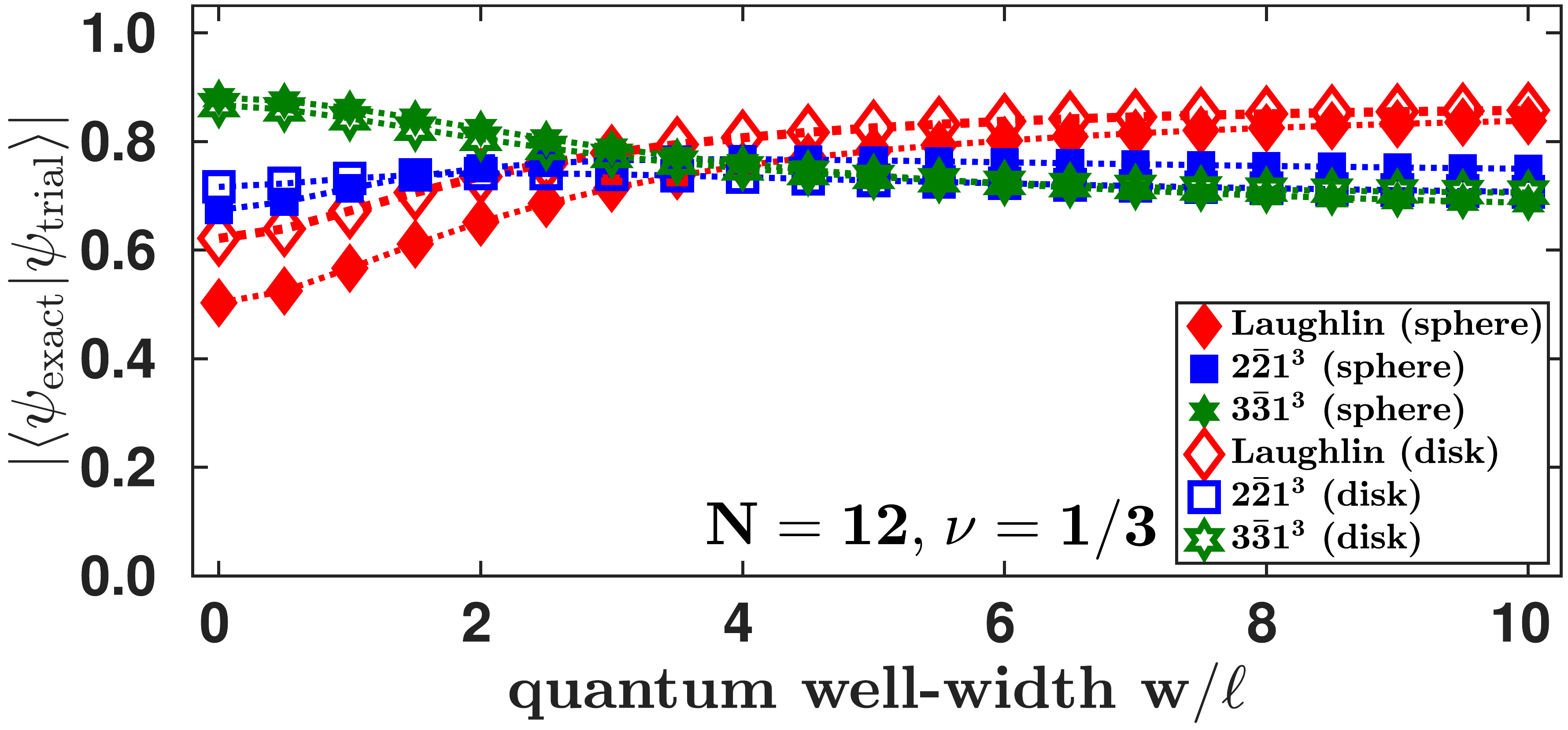} 
\caption{(color online) Overlaps of the $3\bar{3}1^{3}$ (green hexagrams), $2\bar{2}1^{3}$ (blue squares) and Laughlin (red diamond) states at $\nu{=}1/3$ with the exact second Landau level Coulomb ground state evaluated in the spherical geometry using the spherical (filled symbols) and disk (open symbols) pseudopotentials for various quantum well-widths $w$ for $N{=}12$ electrons. }
\label{fig: overlaps_1_3_SLL_3bar3111_2bar2111_Laughlin}
\end{center}
\end{figure}

In Ref.~\cite{Faugno20b}, it was also proposed that the transitions between these Abelian states can be induced in the SLL by varying the quantum well-width. For completeness, we present results here that support this proposal.  To take into account the effect of the finite thickness of the quantum well, we consider an interaction in which the transverse wave function is modeled as the ground state of a free particle in an infinite square well of width $w$~\cite{Balram20b}. In Fig.~\ref{fig: overlaps_1_3_SLL_3bar3111_2bar2111_Laughlin} we show the overlap of the exact SLL Coulomb ground states obtained using the spherical and disk pseudopotentials with the Laughlin, $2\bar{2}1^{3}$ and $3\bar{3}1^{3}$ state for a system of $N{=}12$ electrons at $2l{=}33$ for quantum well-widths $w{\leq}10\ell$.  We find that the $\mathbb{Z}_{2}$ and $\mathbb{Z}_{3}$ states have a higher overlap than the Laughlin states at small widths while the Laughlin state has a higher overlap than the $\mathbb{Z}_{2}$ and $\mathbb{Z}_{3}$ states at larger widths. This result supports the assertion made in Ref.~\cite{Faugno20b} that the Laughlin state at larger widths could transition into a $\mathbb{Z}_{n}$ state at smaller widths.  The value of the width at which this transition takes place was estimated to be $w_{c}{\sim}1.5\ell$~\cite{Faugno20b} and is roughly consistent with the width in the vicinity of which for the system of $N{=}12$ electrons the overlap of the Laughlin state drops and the $\mathbb{Z}_{2}$ and $\mathbb{Z}_{3}$ states pick up as the width is reduced (see Fig.~\ref{fig: overlaps_1_3_SLL_3bar3111_2bar2111_Laughlin}). The charge and neutral gaps for the $7/3$ state as a function of the width were already shown in Ref.~\cite{Balram20b}. The results there suggest that the FQHE at $7/3$ remains robust for all widths in the range $[0,10\ell]$.

\twocolumngrid	

\bibliography{biblio_fqhe}
\bibliographystyle{apsrev}
\end{document}